\documentclass[twocolumn,pre]{revtex4-1}
\usepackage{graphicx}
\usepackage{amsmath}
\usepackage{appendix}

\newcommand{\be}{\begin{equation}} 
\newcommand{\ee}{\end{equation}} 
\newcommand{\bea}{\begin{eqnarray}} 
\newcommand{\eea}{\end{eqnarray}} 
\newcommand{\bc}{\begin{center}} 
\newcommand{\ec}{\end{center}}

\begin{document}

\title{Motif statistics of artificially evolved and biological networks}

\author{Bur\c cin Danac\i $^1$, Mehmet Ali An\i l$^2$ and Ay\c se Erzan$^1$}

\address{$^1$Department of Physics Engineering,Istanbul Technical University, 
Maslak, Istanbul, Turkey\\
$^2$Department of Physics, University of Colorado,
Boulder, Colorado, USA
}

\begin{abstract}

Topological features of gene regulatory networks can be successfully reproduced by a model population evolving under selection for short dynamical attractors.  The evolved population of networks exhibit  motif statistics, summarized by significance profiles, which  closely match  those of {\it  E. coli, S. cerevsiae} and {\it B. subtilis}, in such features as the  excess of linear motifs and feed-forward loops, and deficiency of feedback loops. The slow relaxation to stasis is a hallmark of a rugged fitness landscape, with independently evolving populations exploring distinct valleys strongly differing in network properties. 

\end{abstract}
\date{\today}
\maketitle


\section{Introduction}

Dynamical  biological networks, such as gene regulatory networks (GRN) have been the subject of intensive study over more than thirty years~\cite{Kauffman,Thomas,Kadanoff1,Kadanoff2,Aldana,Drossel,balcan,Cheng}. It is generally agreed that topology is the main determinant of the dynamical behavior.  Milo et al.~\cite{alon1} have introduced the useful concept of network motifs, and shown the predominance of feed-forward loops in biological contexts. The significance profiles of the motif composition of several  transcriptional GRNs reveal a marked enhancement of  feed-forward loops and an equally marked absence of feed-back loops~\cite{alon2} (also see Klemm and Bornholdt~\cite{Klemm}.)

The viability or usefulness of biologically meaningful regulatory networks  usually require the dynamics to have a point attractor~\cite{Ciliberti,Ciliberti2}, or a small number of multistationary states~\cite{Thomas,Thomas1,Thomas2,Thomas3,multistat}. This is easy to understand e.g., in the context of tissue differentiation, where an embryonic cell is once and for all committed to a particular tissue type, or  within the context of periodic behavior such as the diurnal cycle.  

In this study, which builds upon and extends previous work~\cite{anil}, we show that simply adopting a fitness function which favors short attractor lengths (point attractors and two-cycles) is sufficient  to evolve, via a genetic algorithm~\cite{geneticalgorithm}, populations of regulatory networks with topological properties found in real-life gene regulatory networks. 
The adjacency matrix of each random graph can be considered as its ``genotype.'' The ``phenotype'' of the network is the set of its dynamical attractors.  The connection between the wiring (the genotype) and the dynamics (the phenotype) is a subtle one, allowing many different types of solutions.  By considering many  different, independently evolving populations, we are able to observe  if interrelations exist between  properties such as the mean degree,  the  motif statistics and the average length and  number of attractors.   

The significance profiles of the motif frequencies of different populations of evolved networks show a marked resemblance to those found by Milo et al.~\cite{alon2} for real gene regulatory networks.  Our results are consistent with earlier findings that the number of feed back loops are suppressed~\cite{Thomas,Klemm} while there is a relatively high frequency of feed-forward loops~\cite{alon1} in biologically relevant regulatory networks.  

The evolutionary paths followed by the different populations  can be very different from each other, nevertheless yielding the desired characteristic of short attractors. This is an indication of a rugged fitness landscape with divergent valleys (where the mean attractor length is minimized), and is reflected in the slow relaxation of the evolutionary process.

In Section 2 we outline our model.  In Section 3 we  present our simulation results.  Section 4 provides a discussion of our findings.

\section{The Model}

Our model networks~\cite{anil} consist of $N$ nodes. Initial populations of random directed networks are generated with an initial  connection probability  $p_0$.  Each element of the initial  adjacency matrix ${\bf A}$,  assumes  values 1 with  probability $p_0$ and 0 with a probability $1-p_0$~\cite{Erdos}.  The directed edges connecting a pair of vertices $(i,j)$ are assigned independently of each other.  Counting each directed edge separately,  the  total number of directed edges is given by $E\,=\, \sum_{ij} A_{ij}$. The in- and out-degrees of each node are initially independent and  distributed according to a Poisson distribution with the mean $p_0N$, and $\langle E \rangle_0 = 2p_0N$.  On the other hand the total connection probability between any two vertices is $p_0(2-p_0)$.  

\begin{figure}[ht]
\begin{center}
\includegraphics[width=11.9cm]{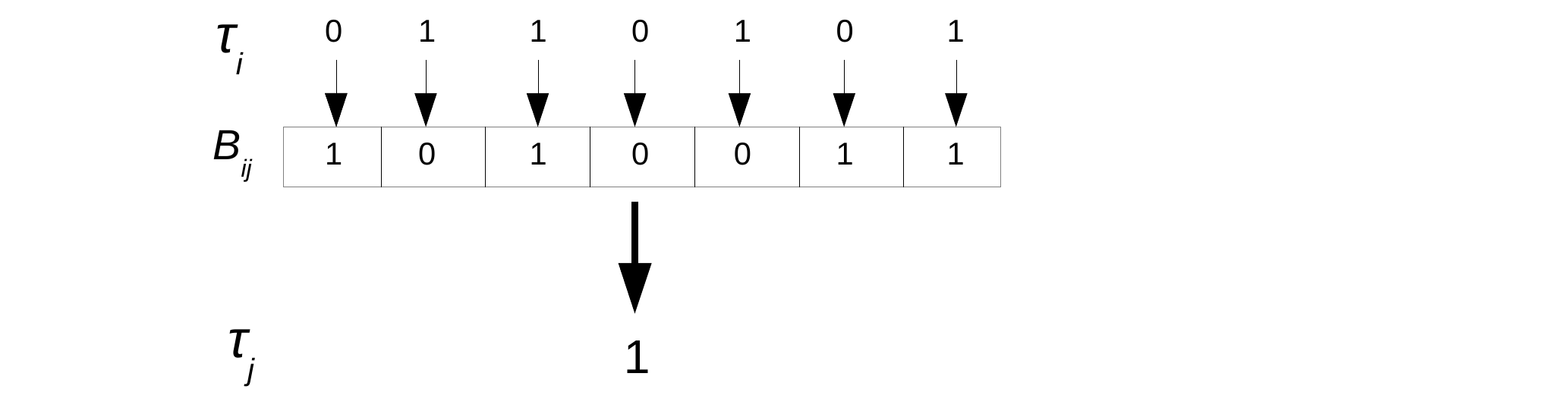}
\end{center}
\caption{\footnotesize  Inputs $\tau_i$ from the neighbors of the $j$th node and the Boolean key ${\bf  B}_j$ determine the output $\tau_j$ according to Eq.(\ref{eq:HS1}).}
\label{fig:bkeys}
\end{figure}

Variables $\tau_i,\; i=1\ldots N$ which can take on the values of 1 or 0, correspond respectively, to an active or a passive state of the node, as in the case of gene regulatory networks.  The state of the system is  given by the vector {\boldmath$\tau$}. 

The dynamics can be specified in two equivalent ways. We assigned a random vector, a  Boolean ``key"  ${\bf  B}_j= (B_{1j}, \ldots,B_{ij}, \ldots, B_{Nj})$ to each $j$th node,  with the entries taking the values 0 and 1 with equal probability.  The nature of the interactions between pairs of nodes are thus predefined and mutations only affect the topology of the graphs by changing the adjacency matrices.  All the networks have the same set of keys associated with their nodes. For each population, the  keys are randomly generated once and for all in the beginning of the simulations.  While we change the wiring of the graph in the course of evolution, the keys have the convenient function of labeling the nodes.

The entries $B_{ij}=0$ or 1, correspond to an activating or suppressing interaction respectively  (see  Table \ref{t:states}).  The synchronous updating is given by a majority rule,
\be \tau_{j} (t+1) = \Theta_{\rm H}\Big( \sum_i^N  A_{ij} \left\{ \left[ \tau_i(t)\; {\rm XOR}\; B_{ij} \right] - \frac{1}{2} \right\}  \Big) \;\;\;.\label{eq:HS1}	
\ee 
The Heaviside step function is defined as  
\be \Theta_{\rm H}(x)= 
\begin{cases} 1 & {\rm for } \;\; x \ge 0 \\ 
0 & {\rm for } \;\; x < 0  \end{cases}.	\label{eq:HS}
\ee 
If there are no incoming edges to the node, $j$,  i.e., or $k_j=\sum_i A_{ij} = 0$,  then $\tau_j(t+1)\equiv\tau_j$. 

It is clear that the two states (active/silent) of a node can just as well be represented by Ising spins $s_i=2\tau_i - 1 =\pm 1$. In this case it is convenient to think of the set of Boolean vectors ${\bf B}_j$ as an  interaction matrix and to define   $\sigma_{ij} =2B_{ij}-1 = \pm 1$, with $\pm 1$ corresponding respectively to an activating or suppressing interaction. The input $s_i=\pm 1$ from the $i$th node to the $j$th node is then processed using $\sigma_{ij}$, and the update rule becomes,  for $k_j\ne 0$,
\be s_{j}(t+1)\equiv  2 \Theta_{\rm H}\left (\sum_i^N A_{ij}\sigma_{ij}s_i(t)\right) -1 \;\;.\label{eq:HS2}
\ee 
The model is then equivalent to a finite  diluted spin glass. This representation also allows us to make direct contact with the work of Thomas and co-workers~\cite{Thomas,Thomas1,Thomas2,Thomas3}.  Note that  an activating (+) interaction means that, if the activating gene is on (i.e., it has the value 1) then it will contribute towards turning the target gene on; conversely, if the activator is off (i.e., has the value 0), this will tend to turn off the target gene.  The complement is true for the repressive (-) interaction; if a repressor is on, this will contribute towards silencing the target gene, but if the repressor is off, then this will contribute towards turning the target gene on. 

\begin{table} [h!] 
\caption{State table for the Boolean keys shown in  Fig.\ref{fig:bkeys}.   The  $B_{ij}$  are the elements of the ``key'' associated with the $j$'th node,  and  $\tau_i$ indicates the state of  the $i$th node.  The output of the $j$th node is  computed via a majority rule ( see Eq.(\ref{eq:HS1} or equivalently, (\ref{eq:HS2})), where we count only the input from nodes connected to $j$ by a directed bond.  This condition is ensured by  the factor $A_{ij}$, which is unity if there exists a bond $(i,j)$ and is zero otherwise. }
\begin{center}
\begin{tabular}[c]{ c c c c c }
\hline\hline
Input\qquad &\quad  Key \quad   &  Output & Interaction &  \\
\hline
$\tau_i$  \qquad  & $B_{ij}$  \qquad &$\tau_i {\rm XOR}\; B_{ij}$ 
&  type & $\sigma_{ij}$ \\
\hline
0   &    0     &        0       &	activating & + \\
1   &    0     &        1       &	activating &+ \\
0   &   1      &		1      &	     repressing & -  \\
1   &   1      &	     0      &	repressing & - \\
\hline\hline
\end{tabular}
\end{center}\label{t:states}
\end{table}

The population of networks is evolved using a genetic algorithm~\cite{geneticalgorithm}. The codes used for the simulations can be accessed from~\cite{kreveik}. We have chosen the fitness function  to depend on the mean attractor length, $ a $, of the network, averaged over the whole phase space, i.e.,  all possible initial conditions, so that each attractor is weighted by the size of its basin of attraction. The fitness function $f( a )$ favors average attractor lengths $a \le 2$. Selected networks are cloned and then mutated by rewiring the edges, while preserving the in- and out-degrees of each node.

The  steps of the genetic algorithm are as follows:

1) Generate a population consisting of randomly wired Boolean graphs, with randomly generated Boolean keys as described above.

2) Select the  graphs to be cloned according to the fitness function $f ( a) = P$ for $  a\le 2$, 0 otherwise. The value of $P=1/2$ was chosen for rapid convergence.

3) Mutate the clones, by randomly choosing two independent pairs of connected nodes and switching the terminals of the two directed edges. This preserves the in- and out-degrees of each node.

4) Remove an equal number of randomly chosen graphs.

5) Go back to step 2.
\begin{figure}[ht]
\raggedleft
\begin{center}
\includegraphics[width=0.5\textwidth]{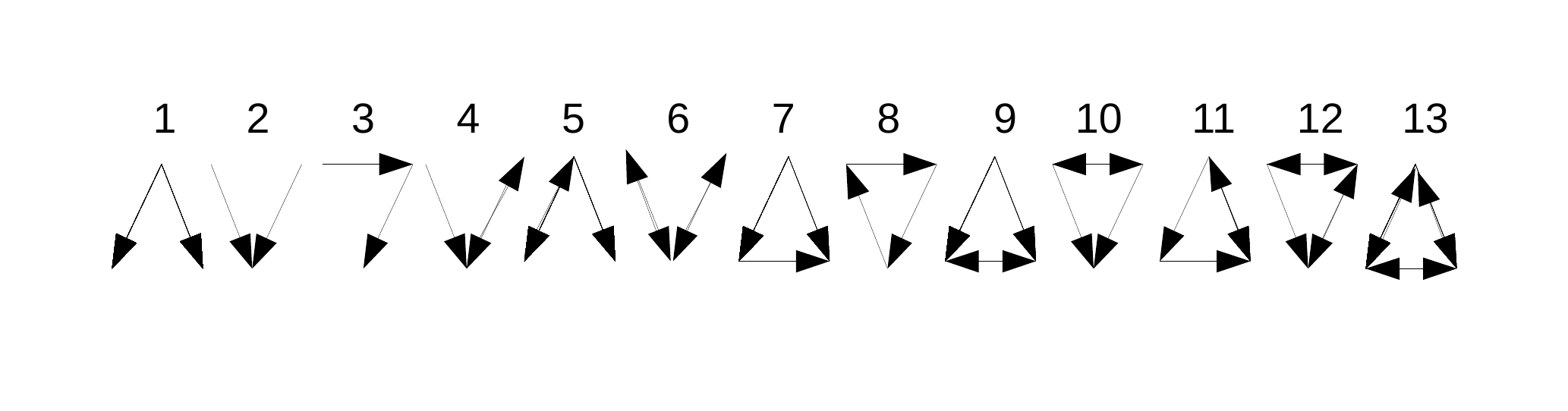}
\end{center}
\caption{\footnotesize Three-motifs without self-interactions. Adapted from~\cite{alon2}.}
\label{fig:mplots}
\end{figure}

In the course of the evolution of the network population, correlations are built up between the edges and the nodes of the networks. 
Some of the higher order features of network topology, beyond single-site properties such as degree
distributions, can be captured  by the frequencies of common motif structures.
 Motifs are subgraphs of networks, consisting of three or more connected nodes~\cite{alon1}. 
Since our networks are small, we considered only 3-motifs and focused our attention on the interactions between the nodes 
by eliminating self-interactions. This yields a total of 13 motifs shown in Fig.~\ref{fig:mplots}. To compare topological features of the resulting graphs to those of 
randomized graphs, $z$-scores and significance profiles based on motif frequencies were computed. 
The $z$-scores are defined as~\cite{alon1},
\be z_\mu = \frac{\langle N_{\rm evol} (\mu) \rangle- \langle N_{\rm rand} (\mu)\rangle}{\sigma [N_{\rm rand}(\mu)]}  	
\label{eq:z-score}\ee 
where $\mu=1,\ldots,13$ is the motif label and $\langle N_{\rm evol} (\mu)\rangle$ and $\langle N_{\rm rand} (\mu)\rangle$ are motif frequencies (evolved and randomized, respectively) averaged over $10^3$ graphs;  $\sigma [N_{\rm rand}(\mu)]$ is the standard deviation.  

Significance profiles, $\mathbf{S}= (S_1, \ldots, S_\mu, \ldots, S_{13})$ for each set of $10^3$ graphs are obtained by normalizing the $z$-scores~\cite{alon2} to give,
\be S_\mu = z_\mu \left(\sum_\mu {z_\mu^2}\right)^{-1/2}\;\;\;.  \label{eq:significance}
\ee
It should be noted that in Ref.~\cite{alon2} the randomization is carried out while keeping the degree sequence fixed, while we only keep the total number of edges fixed,  due to the randomizability problem we encounter with small networks, as explained in the next section.  

\begin{figure}[ht]
\includegraphics[width=7.9cm]{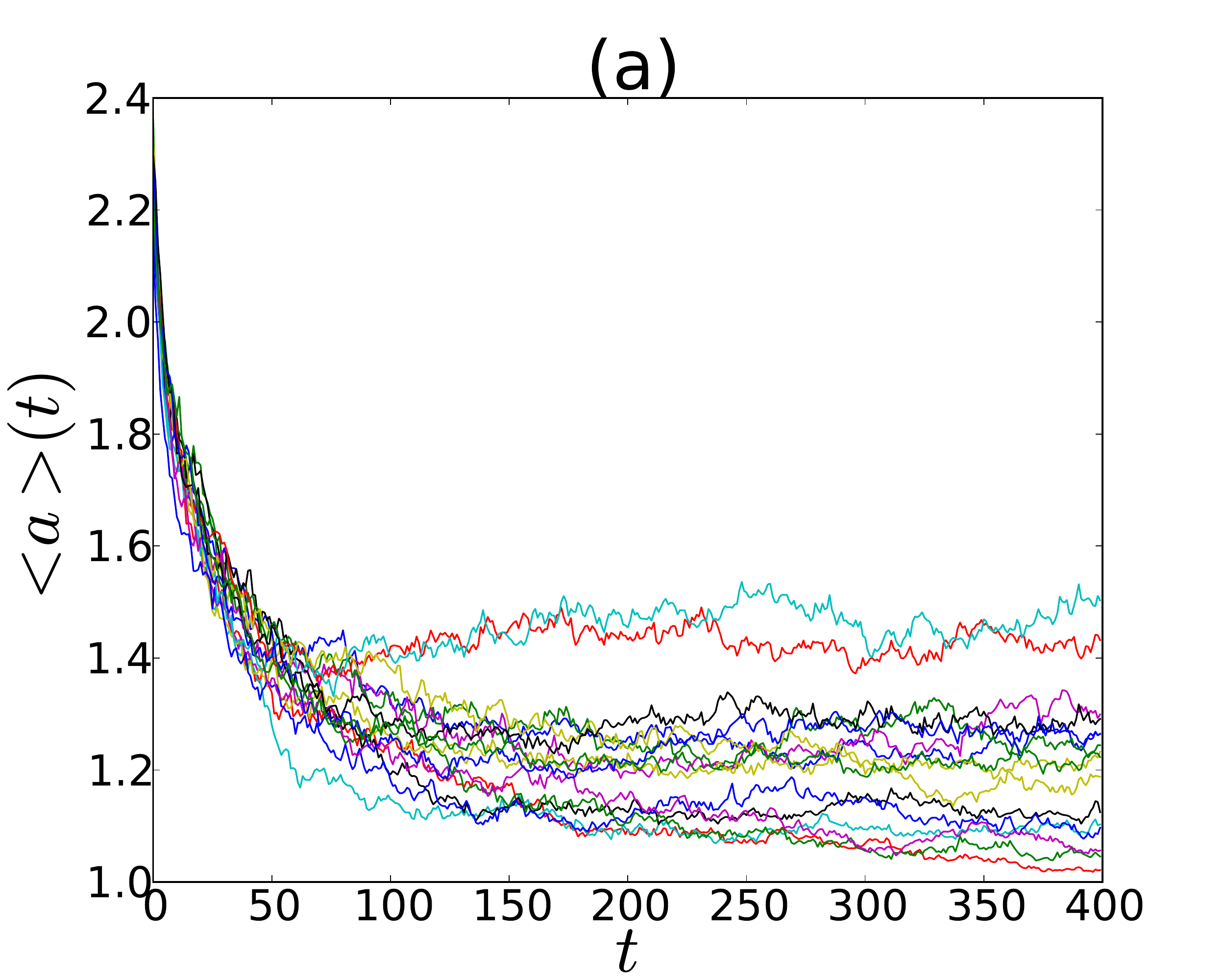}
\includegraphics[width=7.9cm]{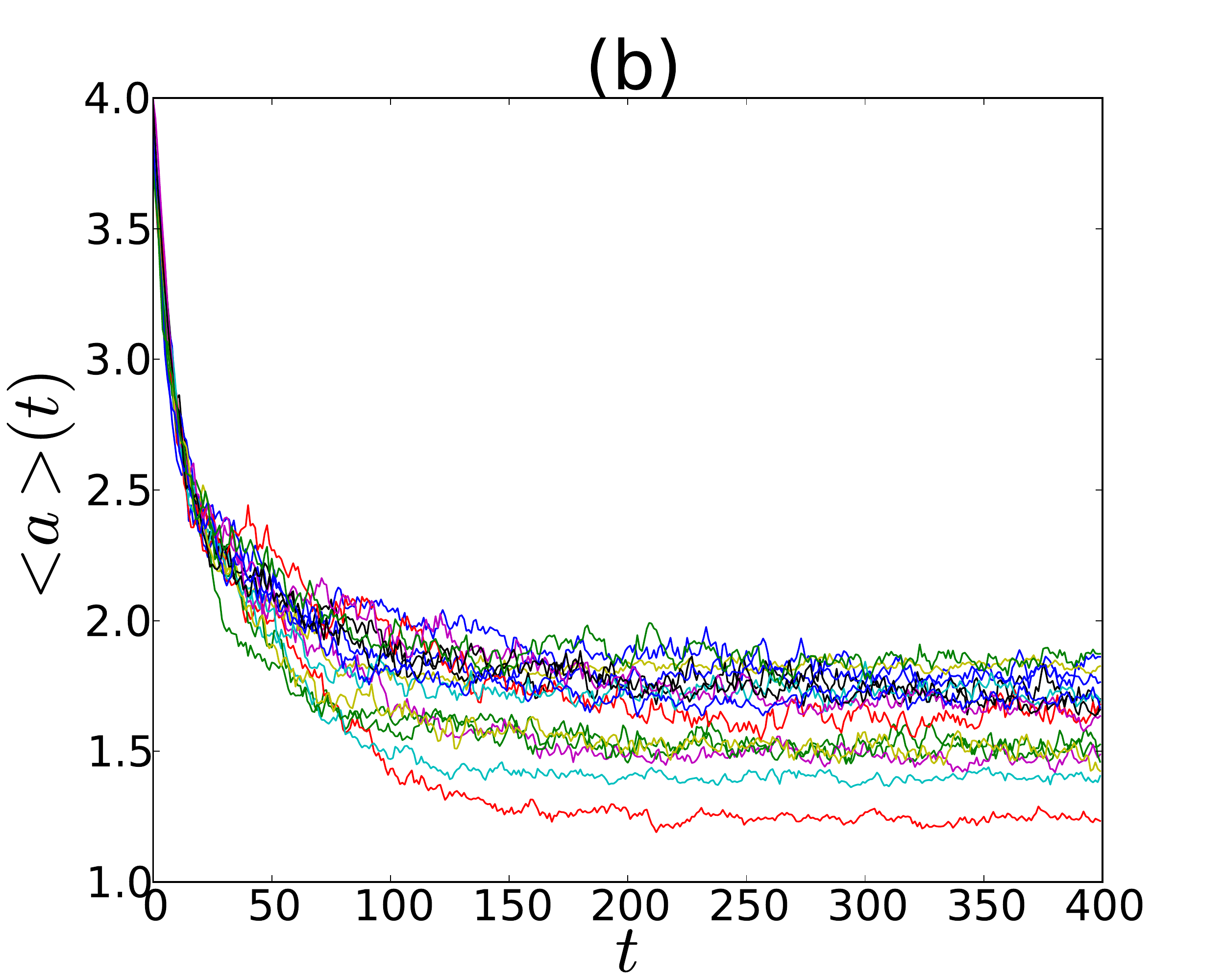}
\caption{\footnotesize  (Color online)  The mean attractor lengths averaged over the whole set, $\langle a \rangle$, v.s. the number of iterations $t$  of the genetic algorithm, for sixteen independent populations with initial connection probabilities (a)  $p_0=0.2$  and  (b) $p_0=0.5$.  The rapid decrease is followed by a slow relaxation region before stasis is reached.  Thereafter, $\langle a \rangle$ for different populations  fluctuates around steady state values. See text.} 
\label{fig:scores1}
\end{figure}

\begin{figure}[ht]
\includegraphics[width=7.9cm]{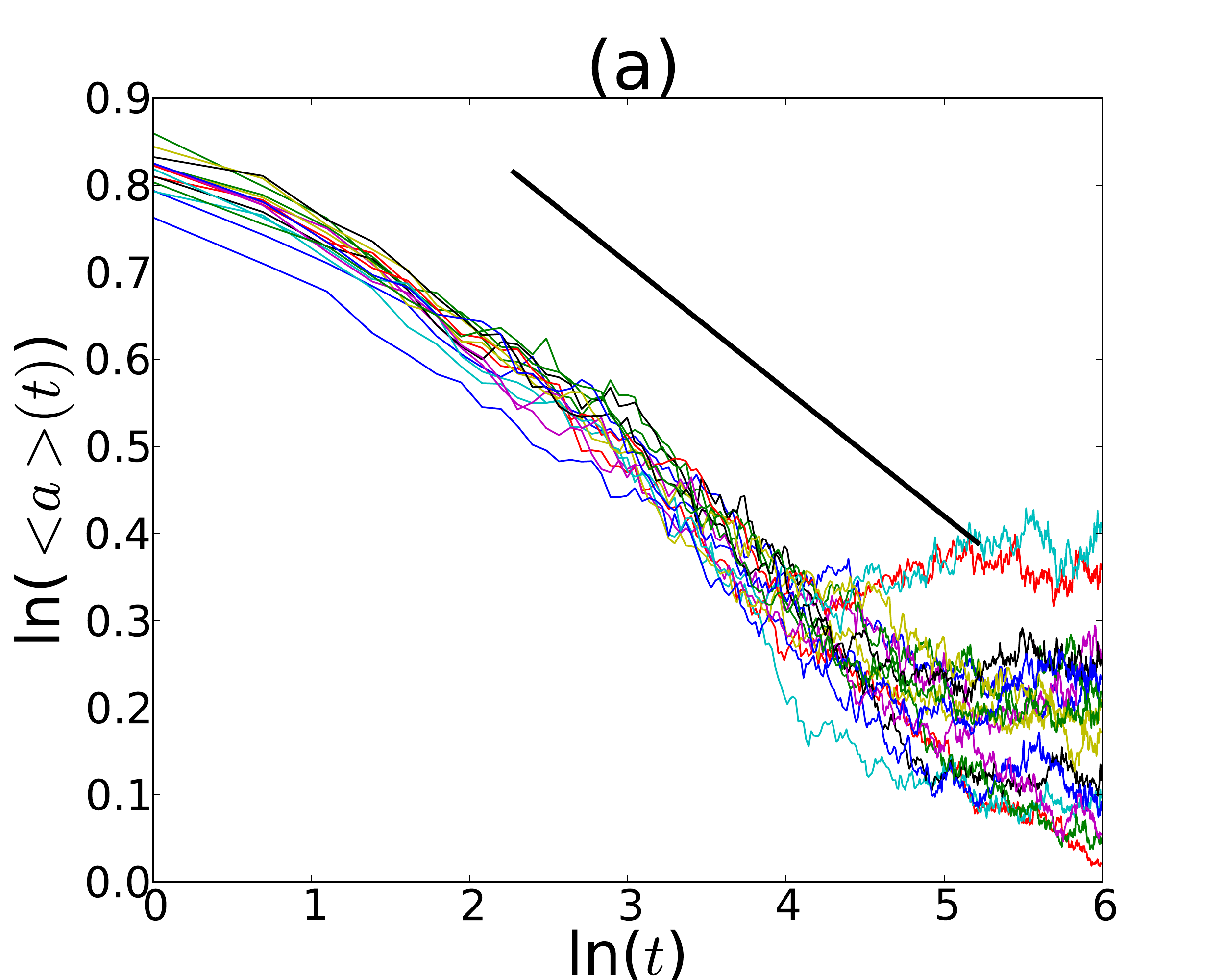}
\includegraphics[width=7.9cm]{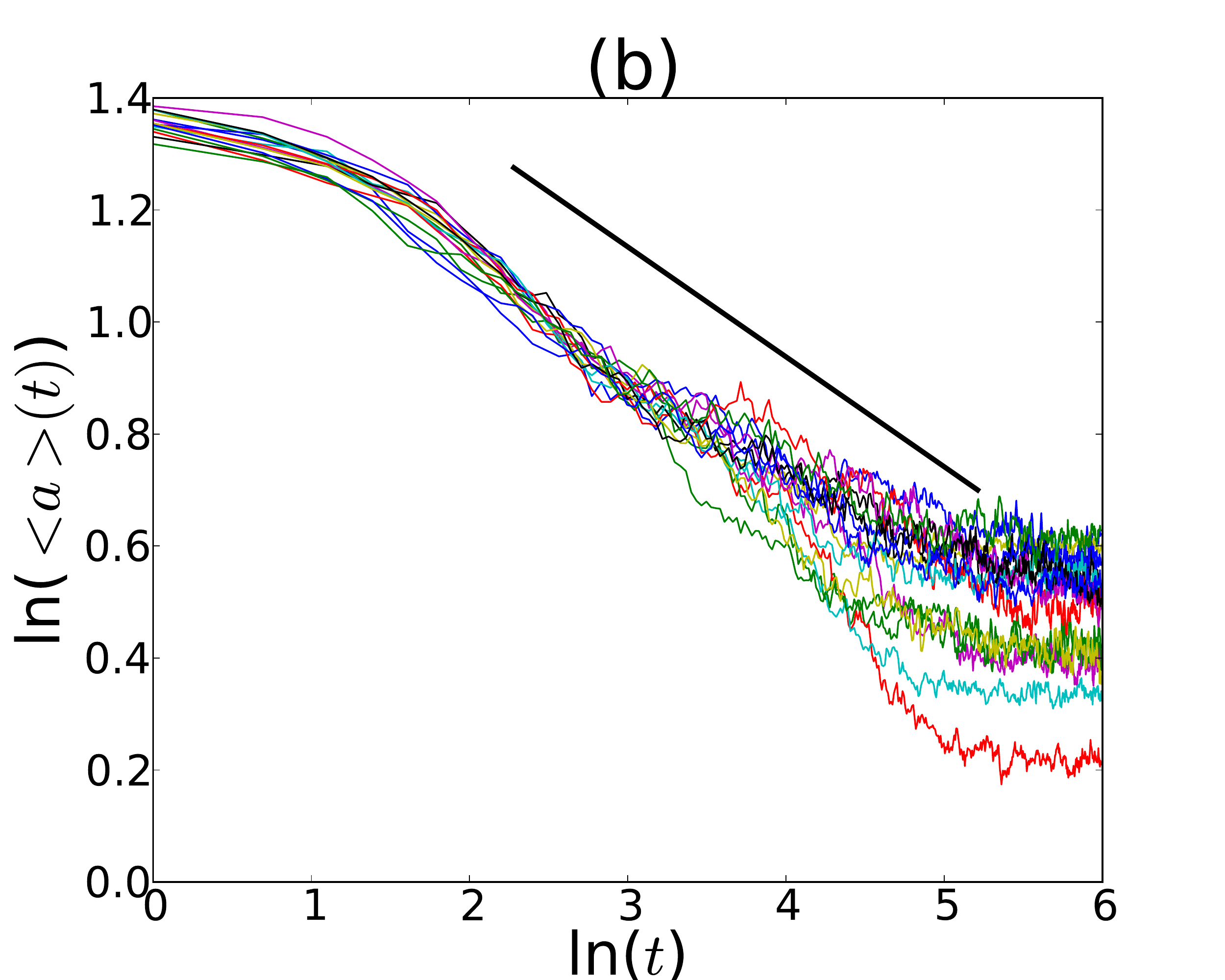}
\caption{\footnotesize  (Color online) Slow relaxation of the mean attractor lengths.  After an initial transient,  $\langle a \rangle$ is seen on this  log-log plot to relax like a power law, $ t^{-\gamma}$, for 16 independent populations with initial connection probabilities  (a)  $p_0 =0.2$  and (b)  $p_0 = 0.5$. The  bold black lines have been obtained by a simultaneous linear least squares fit  to  16 curves over the interval $7<t<150$ and shifted upwards for better visibility.  We find the exponents   $\gamma=0.15 \pm 0.03$  and $\gamma = 0.20 \pm 0.05$  for (a) and (b)  respectively.} 
\label{fig:scores}
\end{figure}

\begin{figure}[ht]
\includegraphics[width=7.9cm]{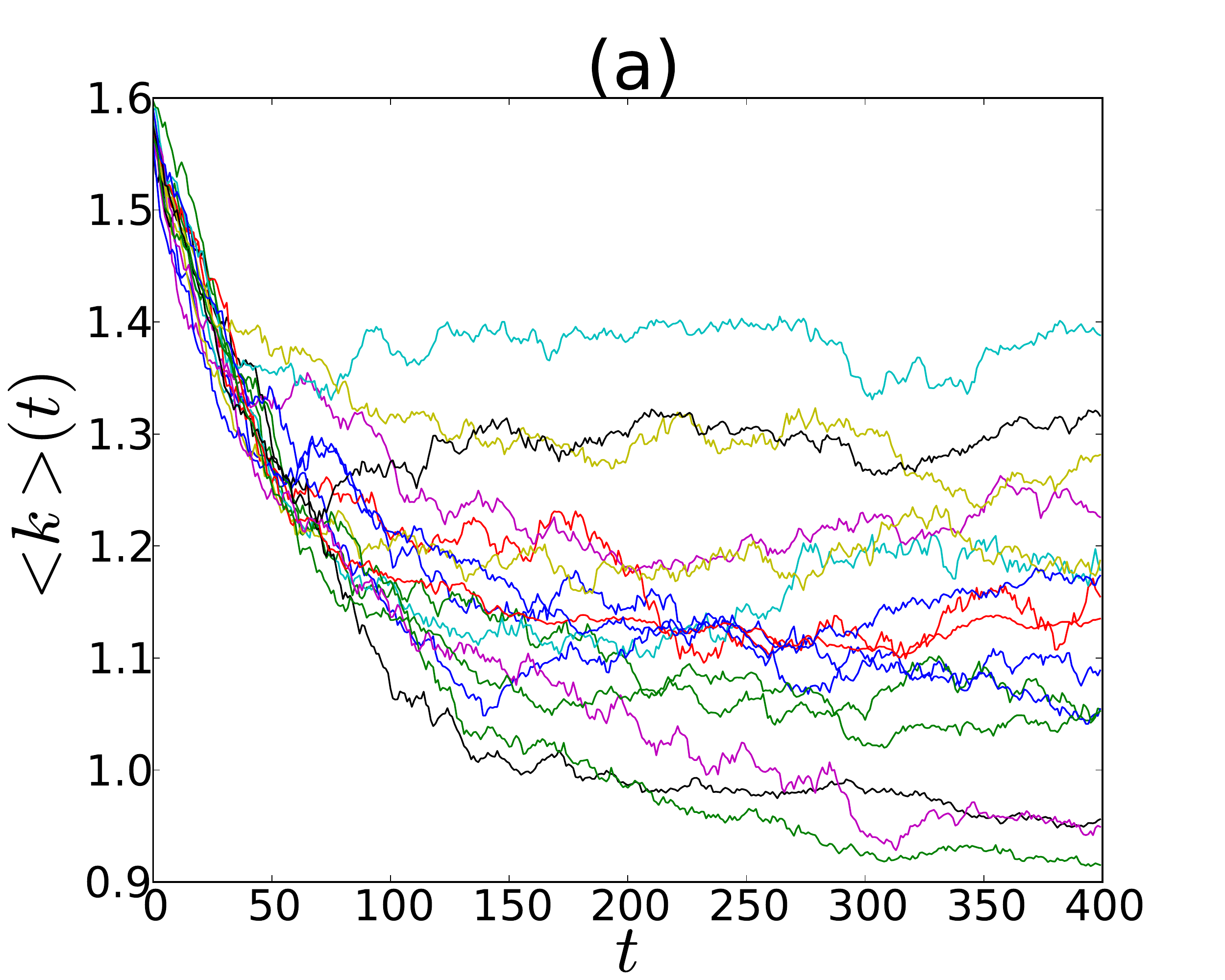}
\includegraphics[width=7.9cm]{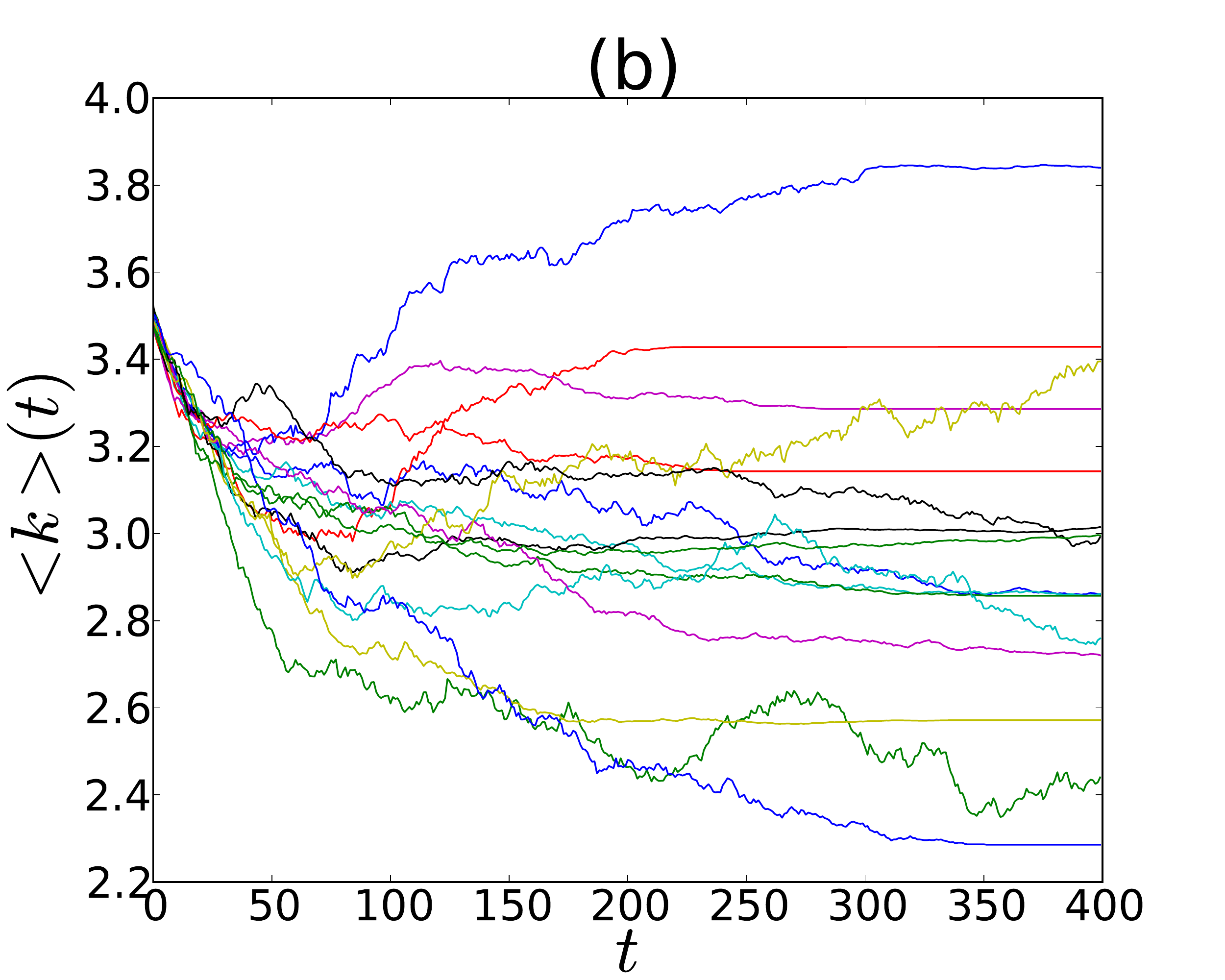}
\caption{\footnotesize  (Color online) The mean degree $\langle k\rangle (t)$ as a function of the number of iterations of the genetic algorithm, for 16 independent populations with initial  connection probability (a)  $p_0 = 0.2$  and  (b) $p_0 = 0.5$.} 
\label{fig:degrees}
\end{figure}

\begin{figure}[ht]
\includegraphics[width=7.9cm]{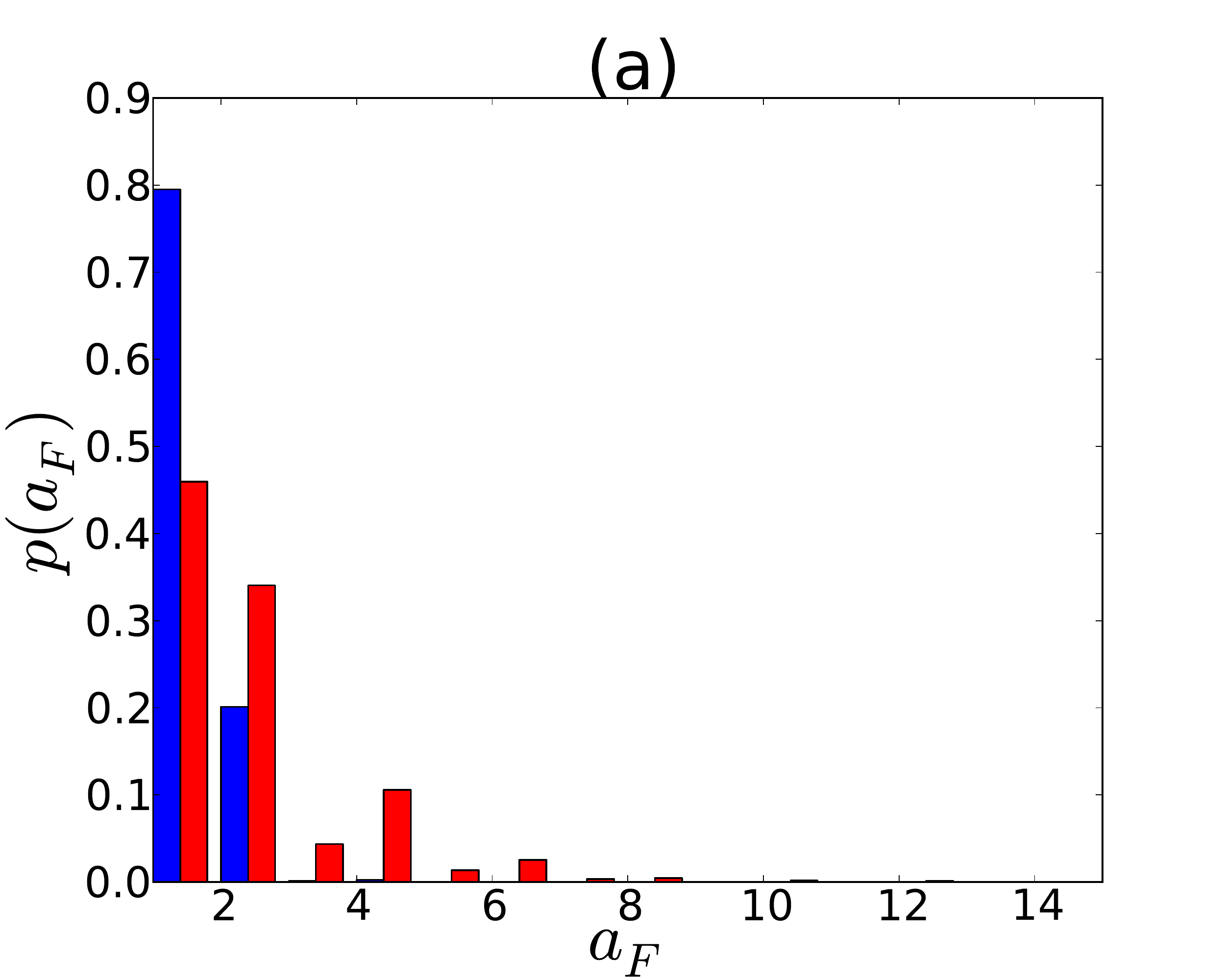}
\includegraphics[width=7.9cm]{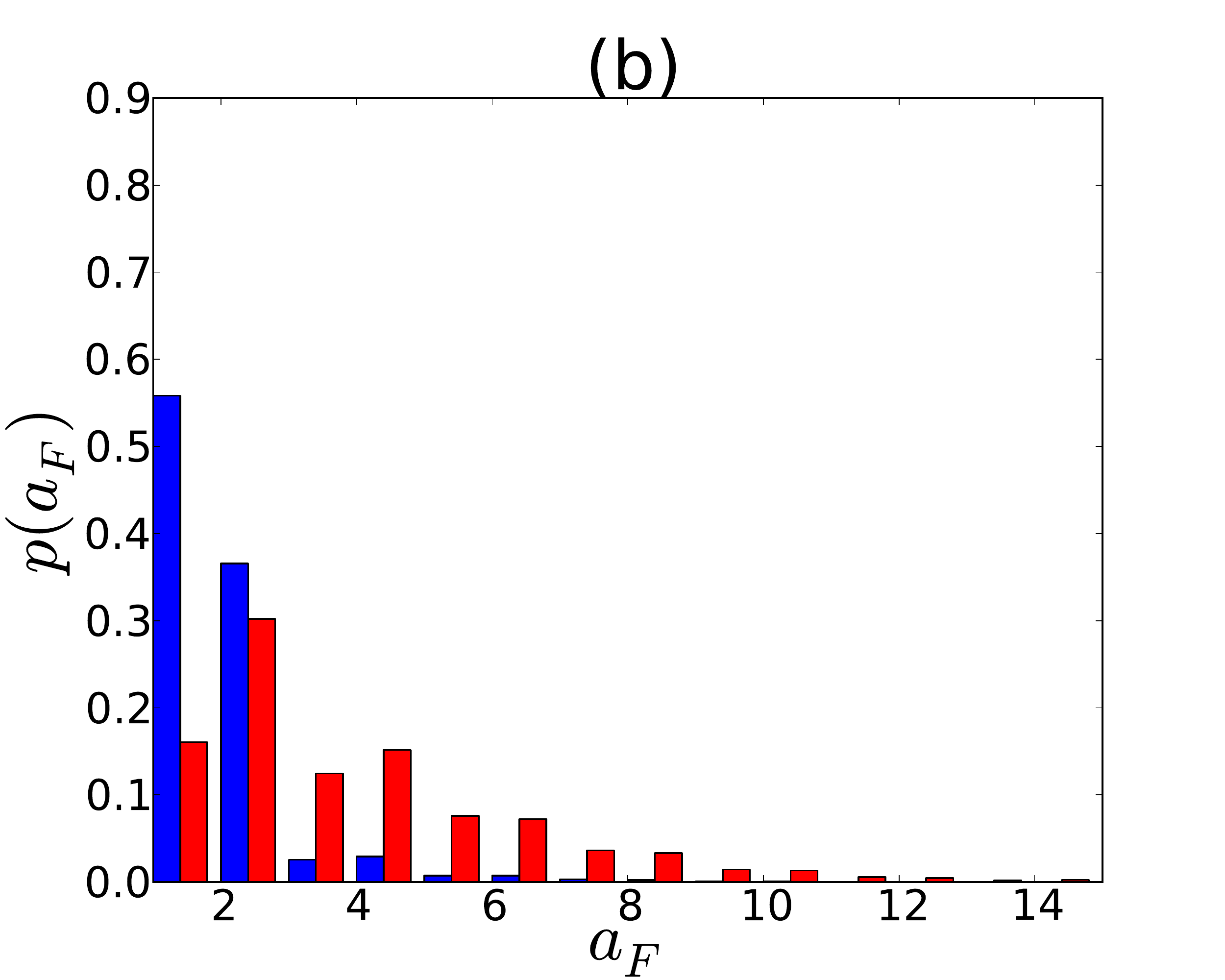}
\caption{\footnotesize  (Color online) Distribution of mean attractor lengths $a_F$, measured at $t=400$ for the evolved (blue) and randomized (red) sets,  with initial  connection probabilities  (a) $p_0 = 0.2$ and (b) $p_0 = 0.5$ . Distributions are computed over $16 \times 10^3$ graphs and all $2^7$ initial states and binned into unit intervals; the bars for the randomized sets are displaced towards the right.   In panel (b),   the distribution of $ a_F$ acquires a longer tail, while the number of attractors decreases (see Fig.~\ref{fig:counts}).} 
\label{fig:score_dist}
\end{figure}

\begin{figure}[ht]
\includegraphics[width=7.9cm]{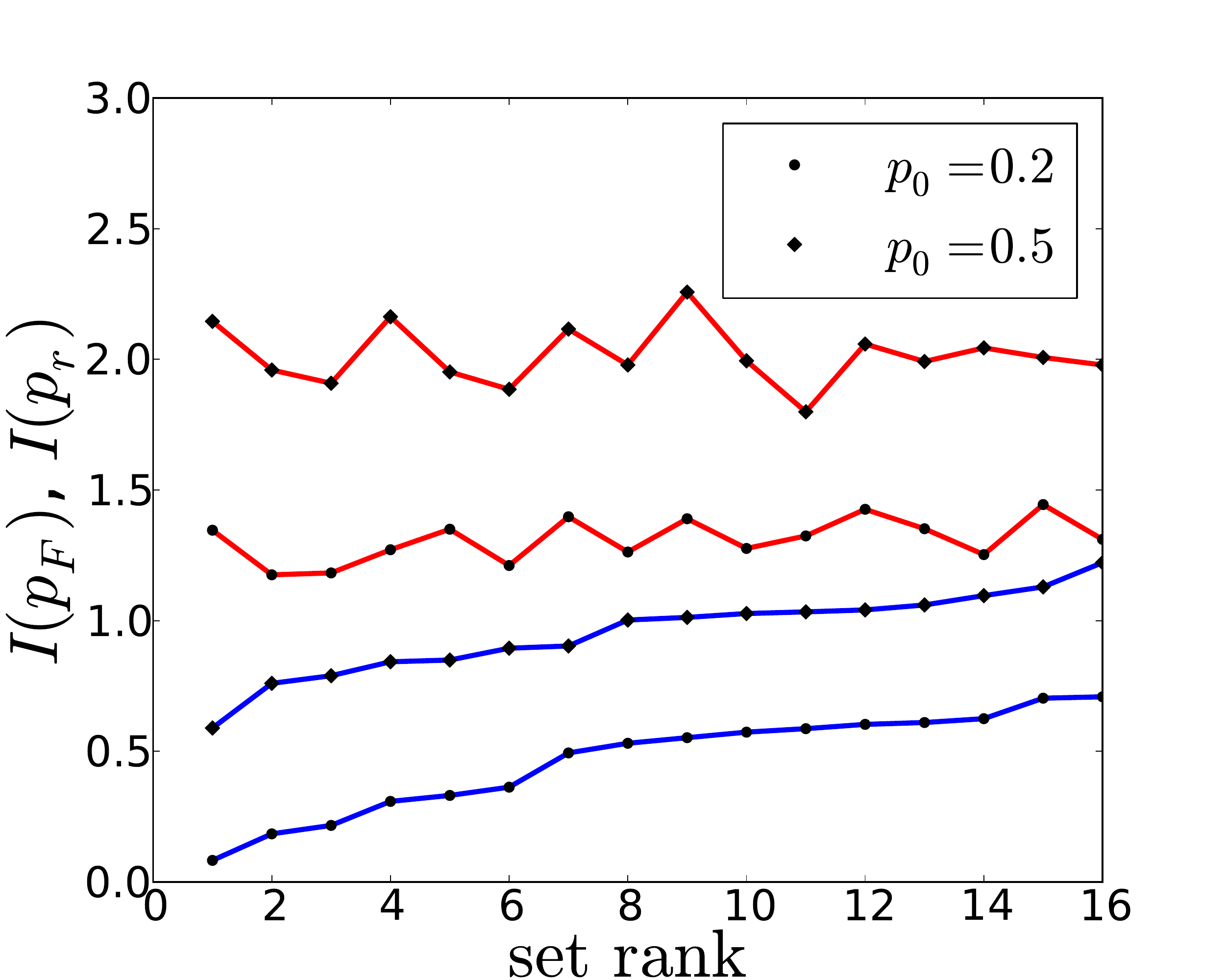}
\caption{\footnotesize (Color online) Information content  $I(p_F)$ of the distributions of the attractor lengths within populations with initial connection probability $p_0 = 0.2$  (circles) and $p_0 = 0.5$ (diamonds), ranked in increasing order of the attractor lengths of the evolved sets. The information content $I(p_r)$  of the randomized distributions (red in color;  first from the top and third from the top) are higher than the corresponding evolved populations (blue in color; second and fourth from the top)  in both cases. The lower information content is a mark of the degree of selection of the evolved population.}
\label{fig:entropies}
\end{figure}

\begin{figure}[ht]
\includegraphics[width=8.1cm]{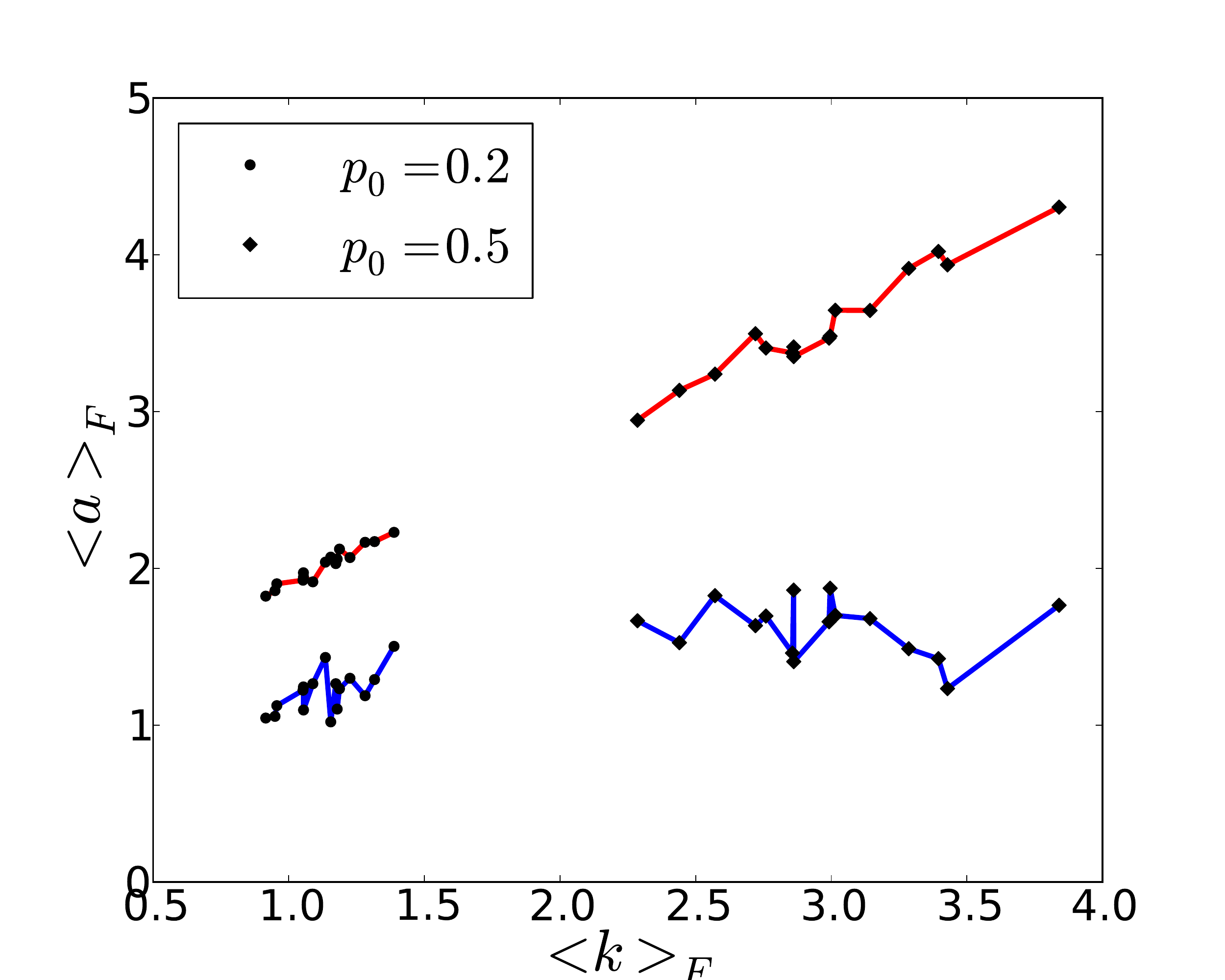}
\caption{\footnotesize  (Color online) Population averages of  mean attractor lengths  $\langle a \rangle_F$ for different evolved populations  (lower graphs, blue in color) and their randomized versions (upper graphs, red in color)   ranked according to their  mean degrees $\langle k \rangle_F$ at $t=400$.  For the evolved networks, there appears to be no correlation between the mean attractor lengths and the mean degrees, while for the randomized networks the two are clearly correlated. } 
\label{fig:scoredegree}
\end{figure}

\begin{figure}[ht]
\includegraphics[width=7.9cm]{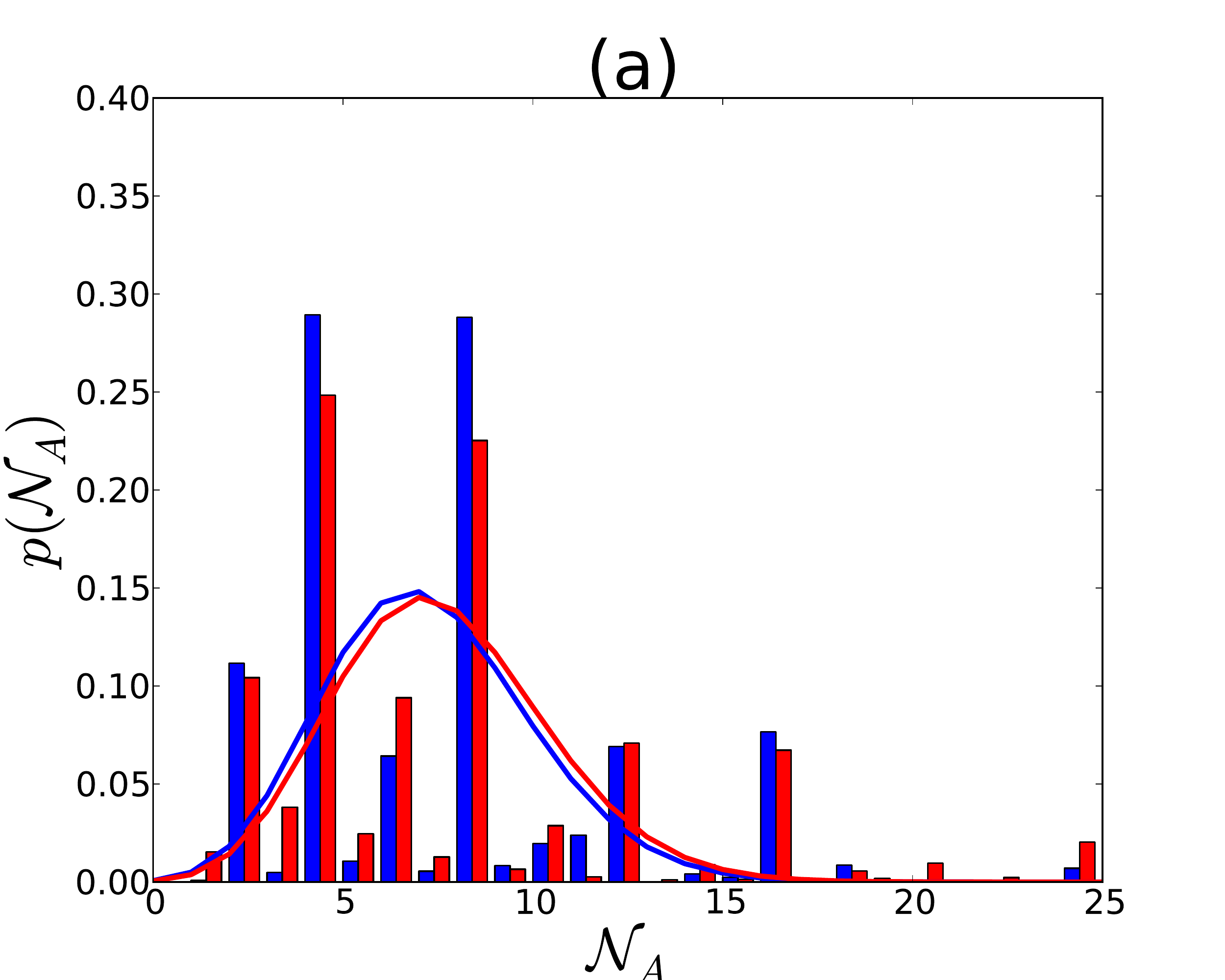}
\includegraphics[width=7.9cm]{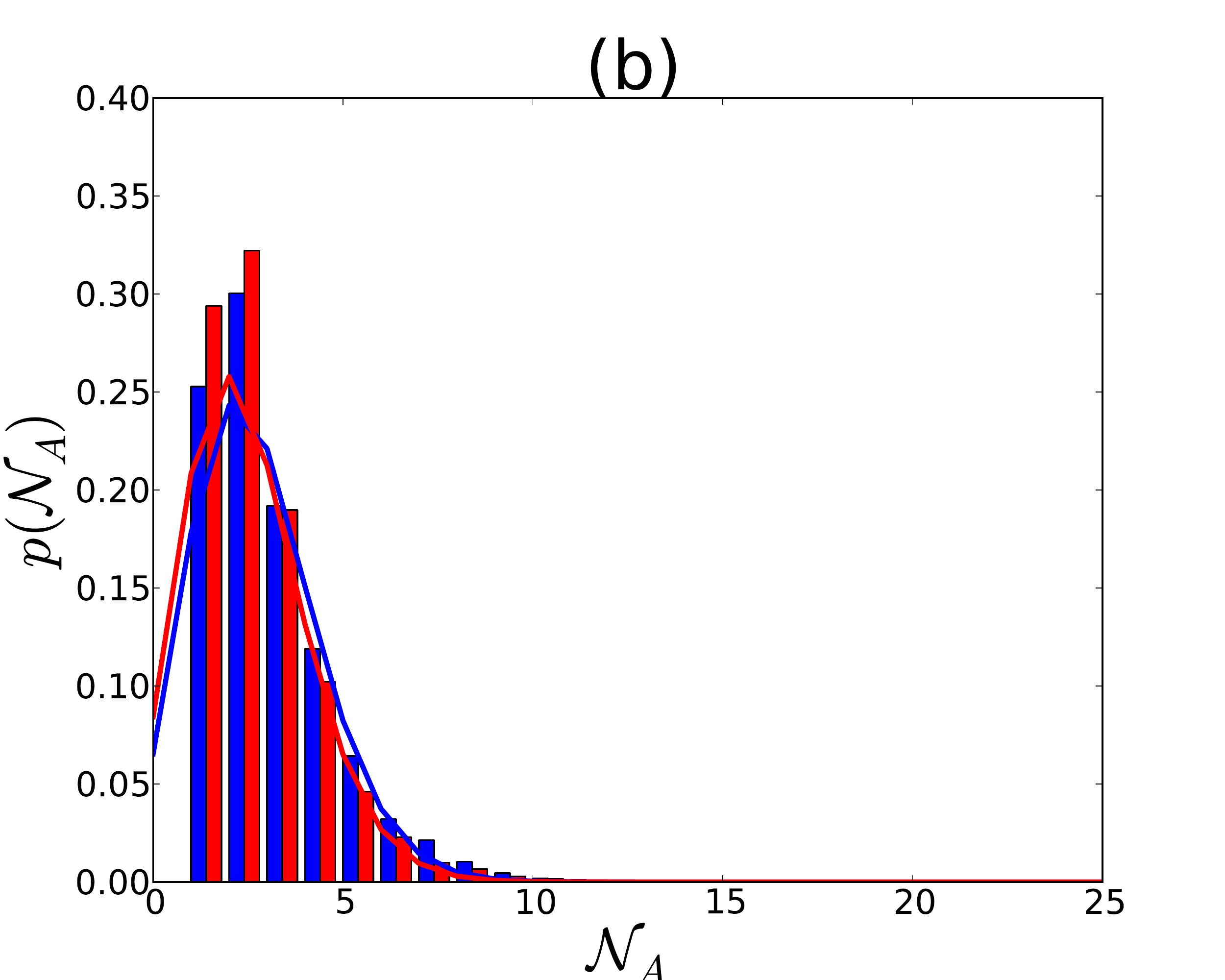}
\caption{\footnotesize  (Color online) The distributions of number of attractors of populations with  (a) $p_0 =0.2$  and  (b) $p_0 = 0.5$. Bars for the evolved and randomized sets are aligned with the integer values (blue in color)  and offset by  0.4 (red in color), respectively.  The blue and red  curves are Poisson distributions with the same mean as the evolved and the randomized graphs, respectively. We see that, the number of different attractors a model graph has, i.e.,   multistationarity, is suppressed as the mean degree is increased. } 
\label{fig:counts}
\end{figure}

\section{Simulations}
\subsection{Simulation procedure}
Our simulations run for small networks with ${N = 7}$, for populations with two different initial connection probabilities $p_0 = 0.2$ and $p_0 = 0.5$. In each case, 16 populations with $10^3$ networks each were generated. Changes in the mean attractor lengths and in the mean degrees during the simulations are shown in Fig~\ref{fig:scores} and in Fig~\ref{fig:degrees}. Graphs with low mean degrees have lower probabilities of being connected. Therefore, for the populations with $p_0 = 0.2$, more graphs were generated and only connected ones were kept at each time step. For the populations with  $p_0 = 0.5$, all the generated graphs  were kept, since only 74 out of sixteen sets of a thousand graphs each ended up being disconnected. 

A randomized control group is used to determine the distinguishing features of the the graphs with short attractor lengths. Randomization was carried out by rewiring the evolved graphs while preserving the total number of edges. Since our networks have relatively few nodes, and since the surviving graphs have typically a smaller density of edges than the initial connection probability $p_0$, the phase space of possible graphs is very small.  This means that rewiring subject to the constraint of preserving the in- and out-degrees very often yields one of the graphs that already belongs to the evolved population, i.e., we have a non-randomizability problem.  For this reason, we required only the total number of edges to be kept constant. 

Preserving  the number of edges constant leads,  once more, to  disconnected networks after randomization, especially for the populations with  $p_0=0.2$. Therefore, more than 60\% of the graphs in the randomized counterparts of the sets with initial $p_0=0.2$ ended up being disconnected, whereas this percentage was less than 1\% for the randomized counterparts of the sets with $p_0=0.5$.

\subsection{Simulation results}

During the simulations, the mean attractor lengths averaged over each independently evolving set of $10^3$ networks are seen to decrease rapidly for all the sets, and stabilize after around 150-200  time steps (see Fig.~\ref{fig:scores1}).  

The slow decay of the mean attractor lengths in the course of the evolution can be seen in Fig.~\ref{fig:scores}.  We find that we can fit the  relaxation curves with  a power law decay,  $\langle a\rangle \sim t^{-\gamma}$,  over the interval $7<t<150$.  The values for the exponents are given in Table I. 

In Fig.~\ref{fig:degrees}, we display  the mean degree $\langle k \rangle(t) $ averaged over each set, plotted against the number of iterations. In the initial stages of their evolution with the genetic algorithm, the populations tend to undergo  large fluctuations in their mean degree before they stabilize in a local minimum of the attractor length. The trajectories shown in Fig.~\ref{fig:degrees} suggest that the fitness landscape is a rugged one, as suggested by the very slow relaxation, with independent populations taking very different evolutionary paths to their respective, relatively well adapted phenotypic distributions.  Moreover, the genotypic features (e.g., the mean degree) can vary quite a bit between different well adapted populations. 

\begin{table} [h!] 
\caption{Parameters for the power law fits to the slow relaxation to stasis, $7<t<150$, and the mean attractor lengths $\langle a\rangle_S$, averaged over a hundred steps within the ``equilibrium'' region, $300<t<400$, for two  initial connection probabilities.  The error bars for $\gamma$ are calculated from the  rms  error of the linear fit. We also provide the average attractor length $\langle a\rangle_F$ taken $t= 400$ and the average attractor length of the randomized networks, $\langle a\rangle_r$.  The averages are  taken over 16 sets of $10^3$ networks each. }
\begin{center}
	\begin{tabular}[c]{l  c c c c c c c }
		\hline\hline
 $p_0$	  \qquad	\qquad  & 	 $\gamma$     	 \qquad     &   $\langle a\rangle_S$ \qquad  &  $\langle a\rangle_F$ \qquad&  $\langle a\rangle_r$ \\
		\hline
$0.2$ 	  \qquad	&      $ 0.15\pm 0.03$   \qquad   &        $1.21 \pm 0.12$  \qquad  &  1.21\qquad & 2.01        \\
$0.5$ \qquad 	&      $ 0.20\pm 0.05$       \qquad       &      $ 1.63  \pm   0.17$ \qquad & 1.61 \qquad& 3.54        \\
			\hline\hline
		\end{tabular}
	\end{center}
 \end{table}

The mean length of the attractors, averaged over all sets and over a time window $300\le t \le 400$ ( $\langle a \rangle_S$) and  $\langle a \rangle_F$, the average taken at $t=400$,  are  given in Table II for the  two different initial connection probabilities, and compared with  $\langle a \rangle_r$, the average attractor length for randomized versions of the evolved networks. (Numerical results  for these quantities computed over individual sets are provided in the Supplementary Materials~\cite{Suppl}.) 

From Table II we see that the all-population averages of the  attractor lengths for the evolved networks are indeed smaller, by almost a factor of two, than the same average over the randomized versions of the evolved networks.  
Even more, striking, however, is the qualitative difference between the distribution  of $a_F$ (blue bars)  and  for the randomized networks (red bars) displayed in Fig.~\ref{fig:score_dist}. The evolved networks have a much narrower and sharply defined attractor length distribution, which we will denote by $p_F(a)$,  compared to their randomized counterparts, which we will denote by $p_r$. 

The narrowing of the distribution of the trait under selection suggests a measure of the response of a population to selection pressure, or in other words, the  {\it selectivity} of an evolutionary process.  In Fig.~\ref{fig:entropies} we compare the  information content (the  Shannon entropy) of the distributions $p_F(a)$ and  $p_r(a)$ for all the different sets.  Defining the selectivity as  the difference between the respective  information contents $I_F$ and $I_r$ and normalized by $I_r$, we have
\be
s\;\equiv\; \frac {I_r -I_F}{I_r} \;\;.
\label{eq:selection}
\ee
We find   $0.46 < s < 0.94$ for $p_0=0.2$ and $0.38 < s < 0.73$ for $p_0=0.5$, with the mean selectivity being $\langle s \rangle = 0.65$ for $p_0=0.2$ and $\langle s \rangle = 0.53$ for $p_0=0.5$.  The smaller selectivity found for $p_0=0.2$  is due to the  small phase space of the networks with sparser edges; we have already remarked upon this non-randomizability problem.

From  Fig.~\ref{fig:scoredegree} we see that, for the evolved sets, there is no correlation between the mean attractor lengths and the mean degrees. On the other hand, for the randomized sets, there is a clear correlation between the mean degree and the mean attractor length; $\langle a \rangle$ increases with increasing mean degrees $\langle k \rangle$, in general.   Aldana et al. have found~\cite{Aldana} a similar result for  Kauffman networks with random  Boolean functions, where  a greater density of edges leads to longer attractors.  

The averaged distribution of the number of attractors ${\cal N_ A}$ of the evolved and randomized graphs is given in Fig.~\ref{fig:counts}.  We observe that for $p_0=0.2$, there is a clear dominance of attractors which are transformed to each other under an exchange of 0 and 1 (equivalently $\pm 1$), manifest in the selection of even numbers  in the distribution.  This is because many nodes do not have any incoming edges, and there are very few with two or more edges incident on them, so there is no frustration. The total number of attractors are much fewer in the $p_0=0.5$ networks. The larger initial mean degree leads to a larger number of instances where an equal number of 1/0 ($\pm 1$) inputs to the nodes lead to a null argument of the Heaviside functions in Eqs.(\ref{eq:HS1},\ref{eq:HS2}), which breaks the symmetry in favor of 1 as the outcome, allowing us to observe odd numbers of attractors of a given length. 

\subsection{Significance profiles}

\begin{figure}[ht]
\begin{center}
\includegraphics[width=7.9cm]{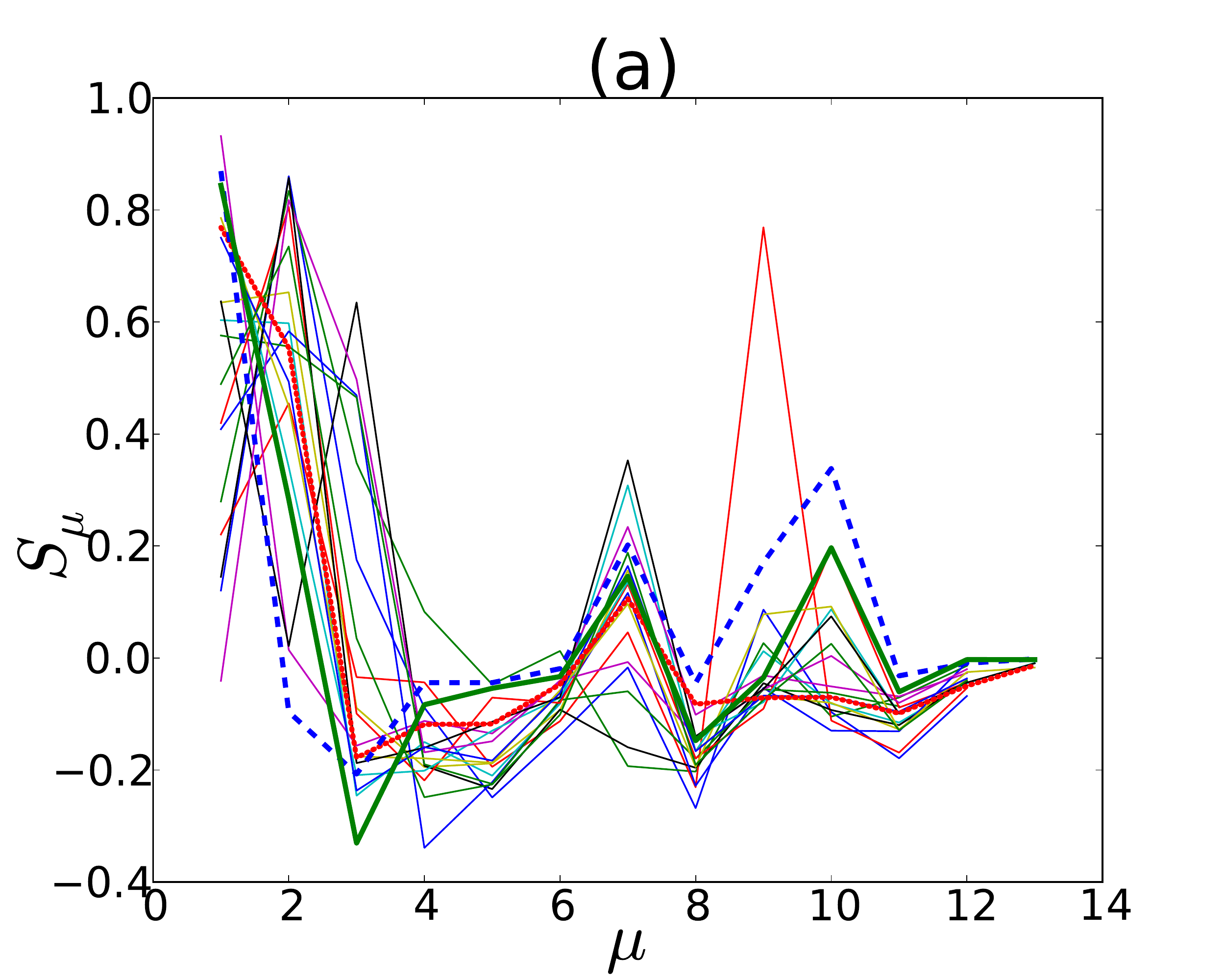}
\includegraphics[width=7.9cm]{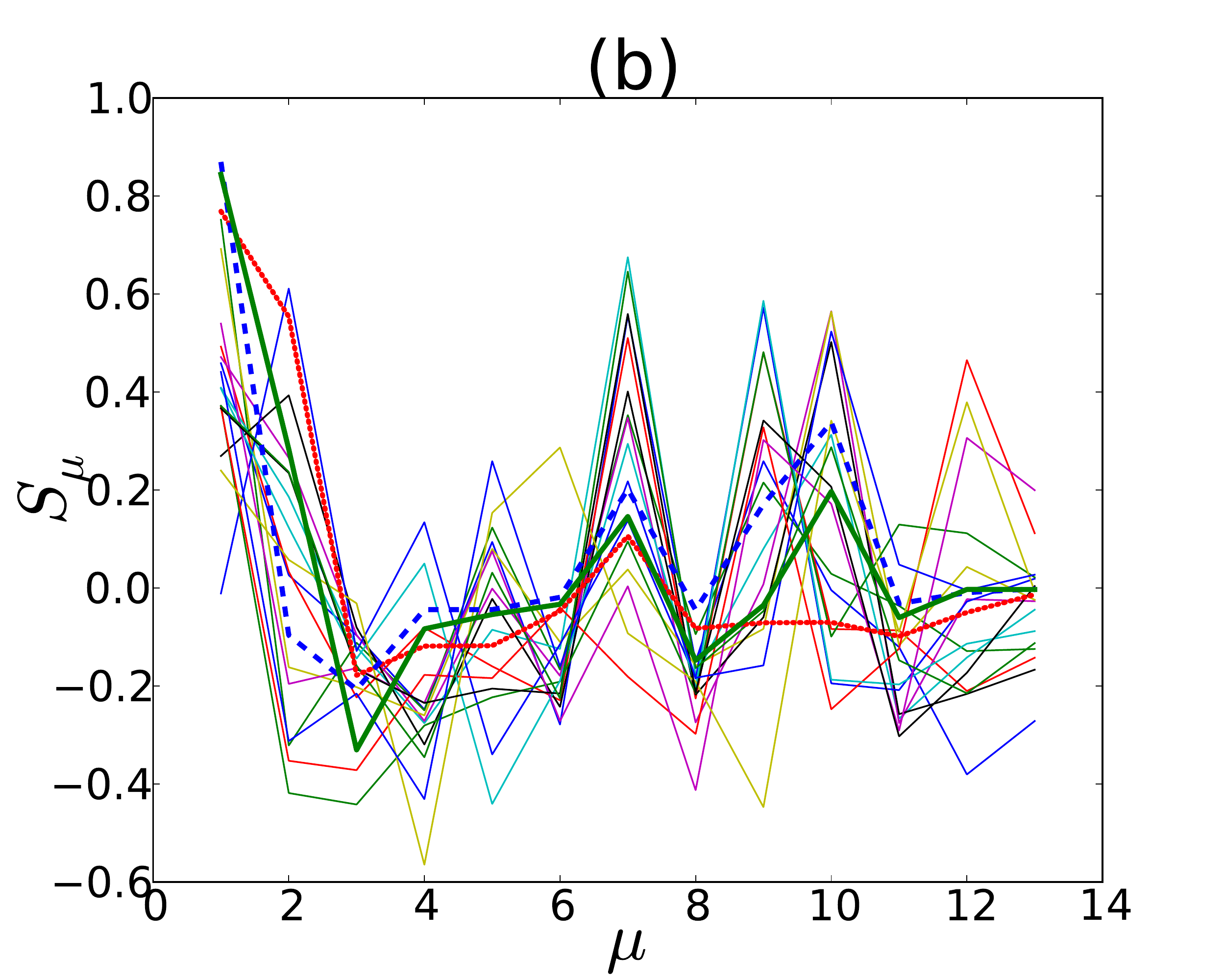}
\end{center}
\caption{\footnotesize  (Color online) Significance profiles $\mathbf{S}$ for populations with initial connection probabilities (a)  $p_0=0.2$ and  (b) $p_0 = 0.5$. The horizontal axis runs over the motif labels, $\mu$, while the vertical axis measures the properly normalized, significant deviation  of motif frequencies from those encountered in the randomized  networks (Eq.\ref{eq:significance}).  Thin lines in different colors have been obtained for  different model populations, while the  bold dotted, dashed and continuous  lines (red, blue and green in color) are the significance profiles of the inner $k$-cores of the gene regulatory networks of  {\it E. coli, B. subtilis} and {\it S. cerevisiae}~\cite{database}, respectively.} 
\label{fig:prof}
\end{figure}

In Fig.~\ref{fig:prof} we display our main results for this paper, the significance profiles (SPs) obtained for the 16 independently evolved populations of $10^3$ model graphs each, for  two different initial connection probabilities.  These two sets of profiles have all been obtained at the 400th  generation of the genetic algorithm, but the profiles show little variation once stasis has been achieved.  The similarity between the significance profiles for the biological networks and the evolved ones is remarkable, since the only selection pressure placed  on the evolving networks was the length of their attractors.   A more detailed discussion is provided in Section 4.

\begin{figure}[ht]
\includegraphics[width=7.9cm]{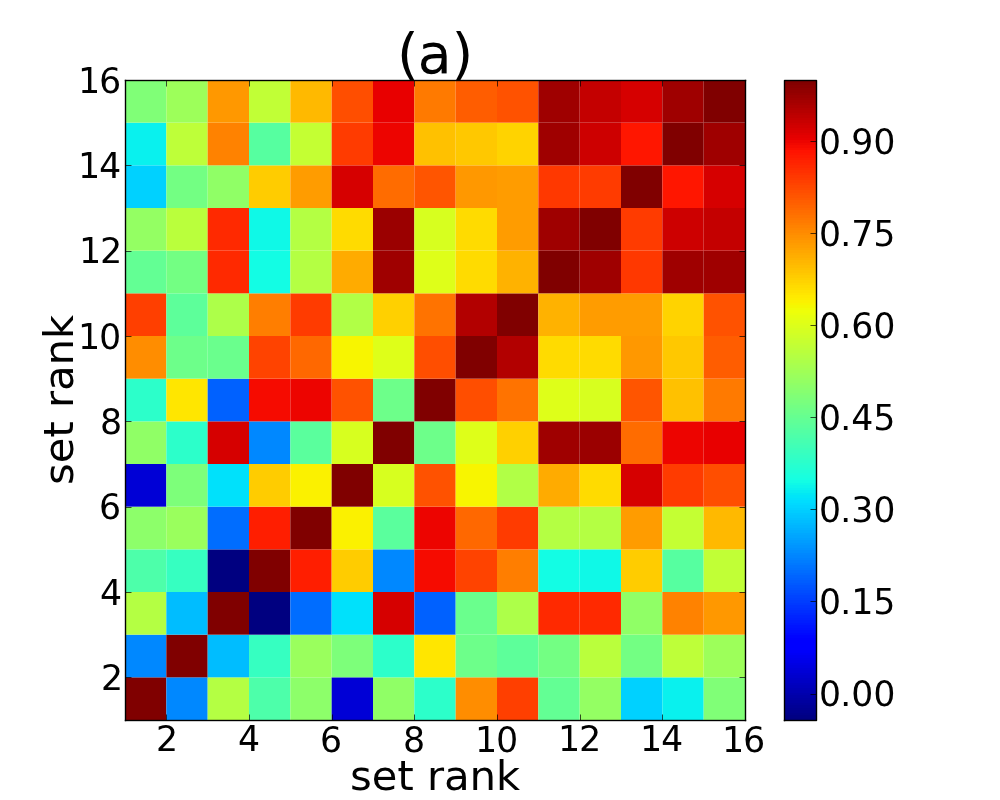}
\includegraphics[width=7.9cm]{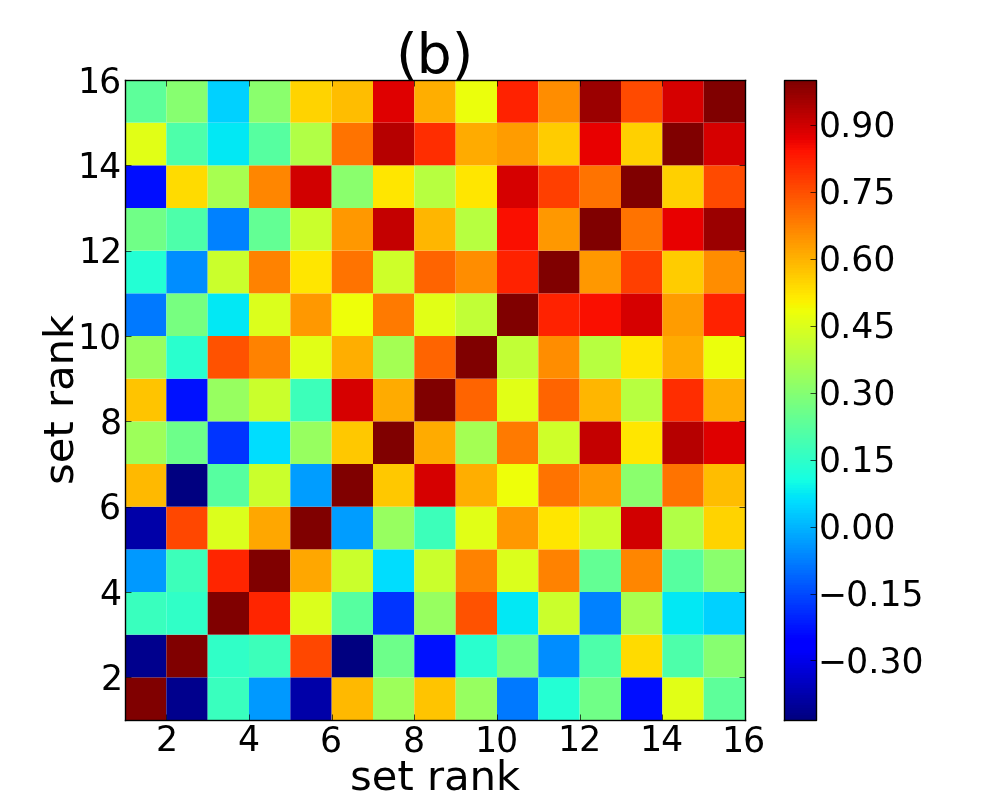}
\caption{\footnotesize (Color online) The overlap between the significance profiles of 16 evolved populations with initial connection probability (a)  $p_0=0.2$  and (b) $p_0=0.5$.  The overlap matrix has been arranged to display blocks of networks with the largest overlaps arrayed along the diagonal, by sorting the sets with respect to the largest total overlap. The numbers correspond to the rank of the set. The color code is given on the side bar.  The overlaps are defined as the scalar product of the normalized SP vectors (see Eq.\ref{eq:overlap}) and take values in the interval $[-1, 1]$.}
\label{fig:O1}
\end{figure}

For a quantitative measure of the similarity  between the significance profiles (Eq.~\ref{eq:significance}) of different sets  $\alpha, \;\beta=1, \ldots M$ of evolved or randomized networks, as well as the overlap of the evolved networks with biological transcriptional gene regulatory networks (TGRNs) we have computed the following scalar product, 
\be
{\cal O}(\mathbf{S}^{(\alpha)},\mathbf{S}^{(\beta)}) = \sum_\mu S^{(\alpha)}_\mu S^{(\beta)} _\mu\;\;.\label{eq:overlap}
\ee
The average overlap of a given set  $\alpha$  with all  the other sets is,
\be
\bar {\cal O}^\alpha = \frac 1 {M-1} \sum_{\beta \ne \alpha}{\cal O}(\mathbf{S}^{(\alpha)},\mathbf{S}^{(\beta)})\;\;.\label{eq:overlapav}
\ee
The results are reported in Fig.\ref{fig:O1} for the evolved sets and in Fig. A.2 for their randomized versions. The sets have been sorted with respect to  $\bar {\cal O}^\alpha$, the sum of all their overlaps. In Table III we give the overlaps between biological TGRNs.
Numerical results for  the  mean overlap   $\bar {\cal O}^\alpha$ within each  set, as well as overlaps of the individuals sets  with biological networks can be found in the Supplementary Material~\cite{Suppl}.

 In Fig. \ref{fig:O1}a,  one may clearly discern a cluster of five sets with large mutual overlaps, followed by another cluster of three,  and in Fig. \ref{fig:O1}b, an initial cluster of six, followed by two smaller clusters of 2.  It should be noted that off-diagonal clusters correspond to groups of sets which share a subset of features in their SPs. In Fig. A.2 in the Appendix,  where the same matrix is constructed for random networks, this pattern is lost completely, and half the networks have negative overlaps (SP vectors pointing in different directions) indicating more dissimilarity than similarity. 

From Table III, we see that the overlap between the TGRNs of different species  can be as high as 91 \%, or as low as 67 \%.  Although a greater number of examples would be needed to start drawing conclusions, it is remarkable that {\it E. coli} is ``farthest'' from {\it B. subtilis}, with {\it S. cerevisiae} placed somewhere in between. Both  {\it E. coli} and {\it B. subtillis} are  bacteria, while {\it S. cerevisiae} (yeast) is a unicellular eukaryote, i.e.,  on that branch of the phylogenetic tree which includes plants and animals.  It seems that this does not make these  two species of bacteria more similar to each other in the significance profiles of the inner core of their TGRNs, than they are to yeast.

\begin{table} [h!] 
\caption{ The overlap between the significance profiles, ${\cal O}(\mathbf{S},\mathbf{ S}^\prime)$, of biological networks.}
\begin{center}
	\begin{tabular}[c]{l  c c c c}
		\hline\hline
     		 	 & 	{\it E. coli}  \qquad \quad  &   {\it S. cerevisiae} \qquad  \qquad  & {\it B. subtilis}  &       \\
		\hline
			{\it E. coli} & 1.00 & 0.91 & 0.67&        \\
			{\it S. cerevisiae} & 0.91 &1.00 & 0.88 &          \\
			{\it B. subtilis} & 0.67 & 0.88 &  1.00 &         \\

			\hline\hline
		\end{tabular}
	\end{center}
 \end{table}

\section{Discussion}

Milo and co-workers have found that a number of organisms exhibit regularities in the relative abundance of different network motifs occurring in their gene regulatory networks.~\cite{alon1,alon2}  In this study we have demonstrated that for a simple model with genes having only two states (on or off) and synchronous updates following a majority rule (Eqs. \ref{eq:HS1}, \ref{eq:HS}, \ref{eq:HS2}), it is possible to artificially evolve populations of model regulatory networks exhibiting similar topological features.  This is done by choosing a fitness function which selects for point attractors or at most period-two cycles in the overall dynamical behavior of the networks.  

Our simulations revealed that evolved populations of networks with point attractors or period-two cycles exhibit higher frequencies for certain motifs (Fig. \ref{fig:mplots}) compared to a set of random networks having the same sizes and number of edges (Fig.~\ref{fig:prof}). These are either loopless motifs such as motifs 1, 2, or involve (one or more) feed-forward loops, such as those numbered 7, 9 and 10 in Fig. 2. On the other hand the motifs 3, 4, 6, 8, 12, and 13 are strongly suppressed in most sets.  The motifs 8 and  11-13 involve feedback loops which are known~\cite{Thomas,Thomas1,Thomas2,Thomas3} to give rise to longer attractors for odd number $n$ of negative interactions and multistationarity for even $n$.

We have also compared our model significance profiles (SPs) with those obtained for real life gene regulatory networks. The empirical networks, which are much larger than our small graphs,  are represented here by their innermost $k$-core~\cite{bollobas,kcore}.   We submit that choosing only the innermost core in a $k$-core decomposition may be seen as a way of scaling down (coarse-graining)  the original network while retaining its most relevant features.

It can be seen from Fig.\ref{fig:prof} that there are some marked features which are shared by almost all the evolved sets and  the core graphs of the transcriptional gene regulatory networks (TGRN)  of {\it E. coli, B. subtilis} and {\it S. cerevisiae}~\cite{database}.  The pronounced peaks at motif No. 1,2,7,9,10 and the deep valleys at No. 3,8, as well as the indifferent showing of the motifs No. 4, 11 and 13 are reproduced, even at an exaggerated rate, by more than two thirds of the evolved sets of regulatory networks, for both the initial connection probabilities of 0.2 and 0.5. 

The motif statistics taken over the biological networks considered here~\cite{database}  are significantly different from those
for just the core graphs. It should be remarked that the core graphs reveal a lot more structure than the significance profiles of the complete gene regulatory networks of E. coli, B. subtilis and S. cerevisiae as reported in~\cite{alon2,Klemm}, and are more closely matched by our model SPs. The significance profiles of these core graphs show greater similarity to those of the TGRN of the higher organisms,
such as D. melanogaster and sea urchin (species unspecifed)~\cite{alon2}, especially in the region of motifs of greater
complexity, (numbers  8-13). It can be argued that these subgraphs, belonging  to the most highly connected, computational core
of the TGRN, correspond to the genes that play the most crucial role in regulation.

\begin{figure}[ht]
\includegraphics[width=7.9cm]{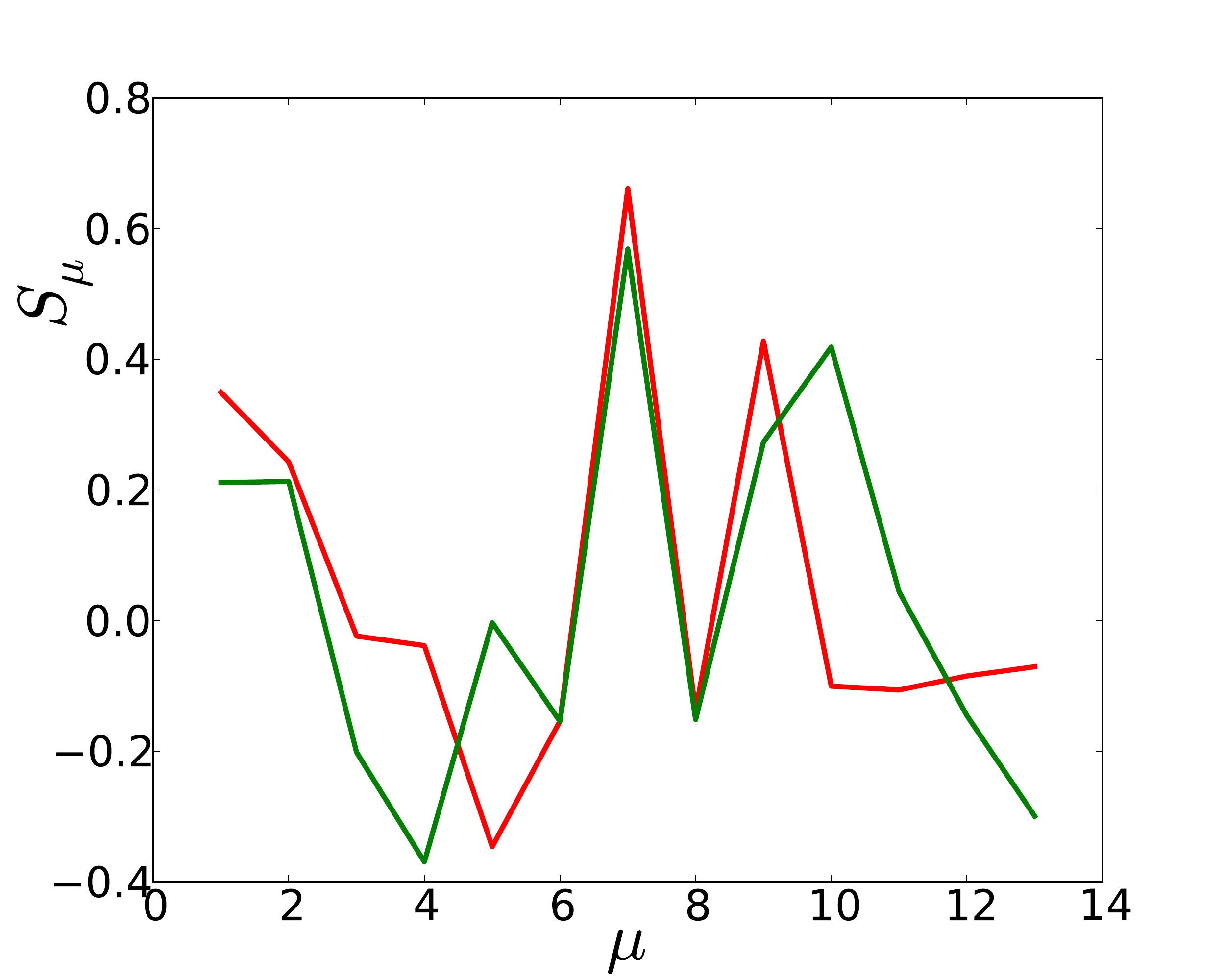}
\caption{\footnotesize  (Color online) Significance profiles of two artificially evolved populations with only activating interactions  (green in color, with the dip at 4 and peak at 10)  and only repressing interactions  (red in color, with the dip at 5 and peak at 9))   with initial connection probability $p_0 = 0.5$.  The characteristic peak at the 7th motif is prominently present in both, while the high profile at 9-11 and  the relatively low profiles exhibited for motifs 4-6 and 8, again resemble the significance profiles in Fig.\ref{fig:prof}.  }
\label{fig:negative}
\end{figure}

Conditions for multistationarity in the asynchronous  dynamics of regulatory networks have been investigated by Thomas and coworkers~\cite{Thomas,Thomas1,Thomas2}, who have found a rule-of-thumb for feedback loops consisting of three nodes.  For asynchronous updates, an even number of ``negative'' (repressive) interactions between the nodes leads to the coexistence   several stable attractors (depending upon the initial conditions),  i.e., multistationarity, whereas  an odd number of negative interactions leads to oscillatory behavior.   For the synchronous updating rules which we have adopted, the presence of an odd number of repressive interaction within a feedback loop leads to the lengthening of the attractors present.   

We find that the dynamics of the Boolean networks which we have studied, at least in so far as they favor point- or at most period-two attractors,  depend much more strongly on the topology of the networks, as characterized by their significance profiles, than the nature (sign) of the interactions.  We have independently evolved sets of  networks where the Boolean keys (Fig.~\ref{fig:bkeys}) were all set to 0 or to 1, leading to uniformly repressive or uniformly attractive interactions.  In comparison to  sets with randomly generated Boolean keys, the attractor lengths were indeed shorter from the outset.  However, the resulting significance profiles shown in Fig.~\ref{fig:negative} exhibit the same structures as in  Fig.\ref{fig:prof}, in particular for the subset of motifs  $\{6,7,8,9\}$ .  The significance profiles at these five motifs can be taken as the topological signature of networks selected for short mean attractor lengths.

{\bf Acknowledgments}
It is a pleasure to acknowledge several useful discussions with Eda Tahir Turan and Neslihan \c Seng{\" o}r.

\begin{center}
{\bf Appendix:}  

{\bf Significance profiles and their overlaps for randomized networks}
\end{center}
\appendix
\setcounter{figure}{0} \renewcommand{\thefigure}{A.\arabic{figure}}
\setcounter{table}{0} \renewcommand{\thetable}{A.\arabic{table}}

\begin{figure}[ht]
\begin{center}
\includegraphics[width=7.9cm]{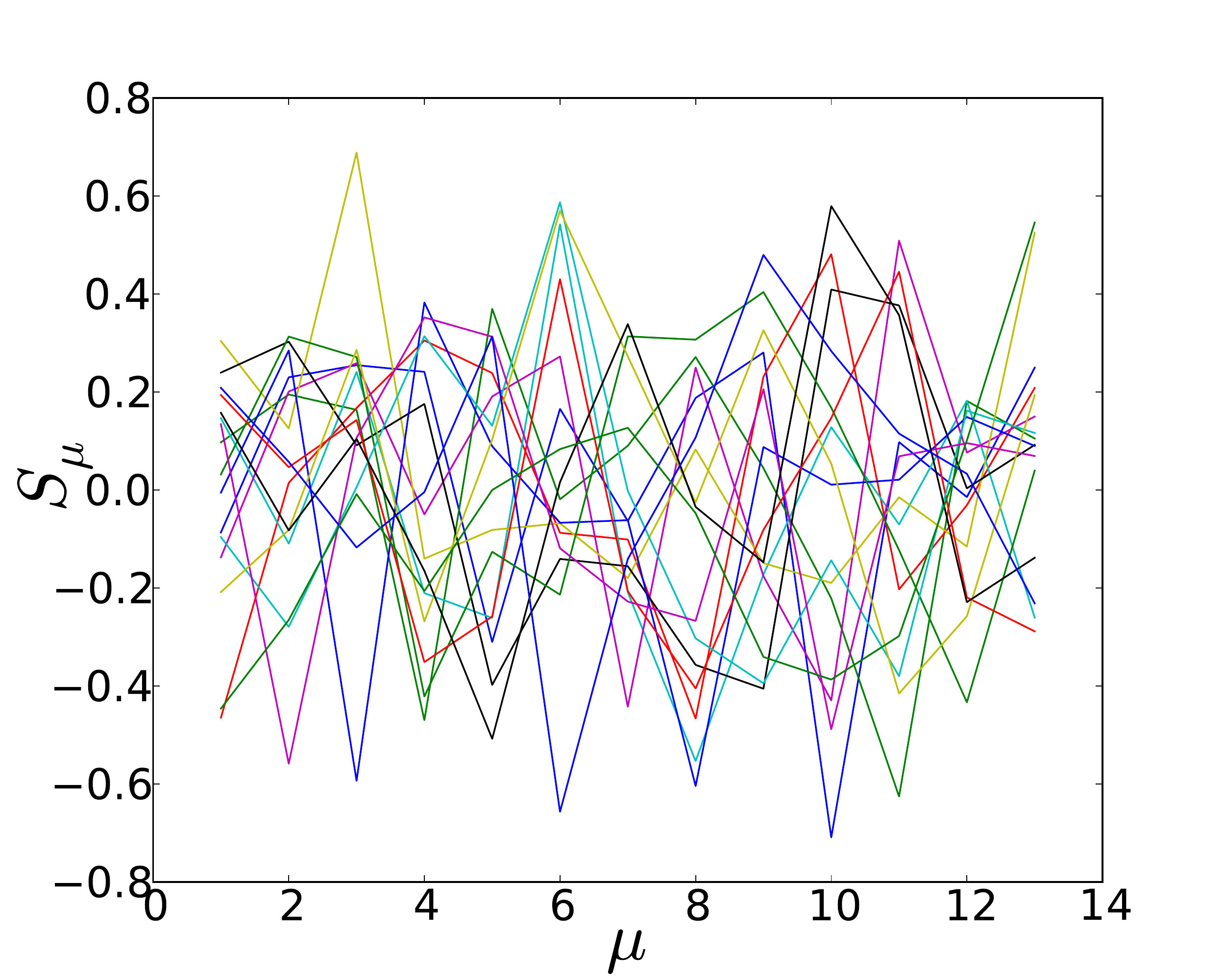}
\end{center}
\caption{\footnotesize (Color online) The significance profiles of 16 randomly generated populations with initial connection probability $ p_0 = 0.5$. No common structure is observed in the profiles.}
\label{fig:random}
\end{figure}

To double check our conclusions regarding the significance profiles (SPs) of evolved sets the,  the $z$-scores and  SPs have been calculated for the null-case, i.e., for  16 sets of $10^3$ randomly generated networks.  As the reference set we have taken an equally large  set of independently generated random graphs. These scores are provided in the Supplementary Material~\cite{Suppl}.   By definition  the expected $z$-scores for a random set are zero for a large enough sample.  Note from (Eq.~\ref{eq:z-score}) that $z_\mu$ has  a $p=0.05$  level of significance only  if $z_\mu >2$ (or p=0.32 for $z_\mu > 1$).  Our $z$-scores  for each  set of random networks are much smaller than one ($-0.10 < z < 0.15$, therefore  without any statistical significance. Moreover  the inter-set standard deviation of the $z$-scores for any given motif, averaged over the 13 motifs is $\bar{\sigma_z}=0.04$, and ranges only from $0.02$ to $0.05$.  The difference between the $z$-scores of the random sets is no more than can be accounted for by  random variation due to under-sampling, as expected. Since  the SPs for each set are scaled  (see Eq.~\ref{eq:significance}) by the standard deviation of the $z$-scores, the profiles for the random sets still show some structure; however there is no coherence between the different random sets, as illustrated in Fig.~\ref{fig:random}. This background is also present in the SPs of the evolved sets, as we explain below. 

The $z$ values for the evolved sets  are much larger in absolute value, the range is $-1.32 < z < 3.17 $; the inter-set standard deviation for different motifs  ranges from $0.18$ to $0.82$, and clearly carries the mark of the convergence as well as the sporadic outliers among the different patterns exhibited in the SPs in Fig.\ref{fig:prof}.   
\begin{figure}[ht]
\begin{center}
\includegraphics[width=7.8cm]{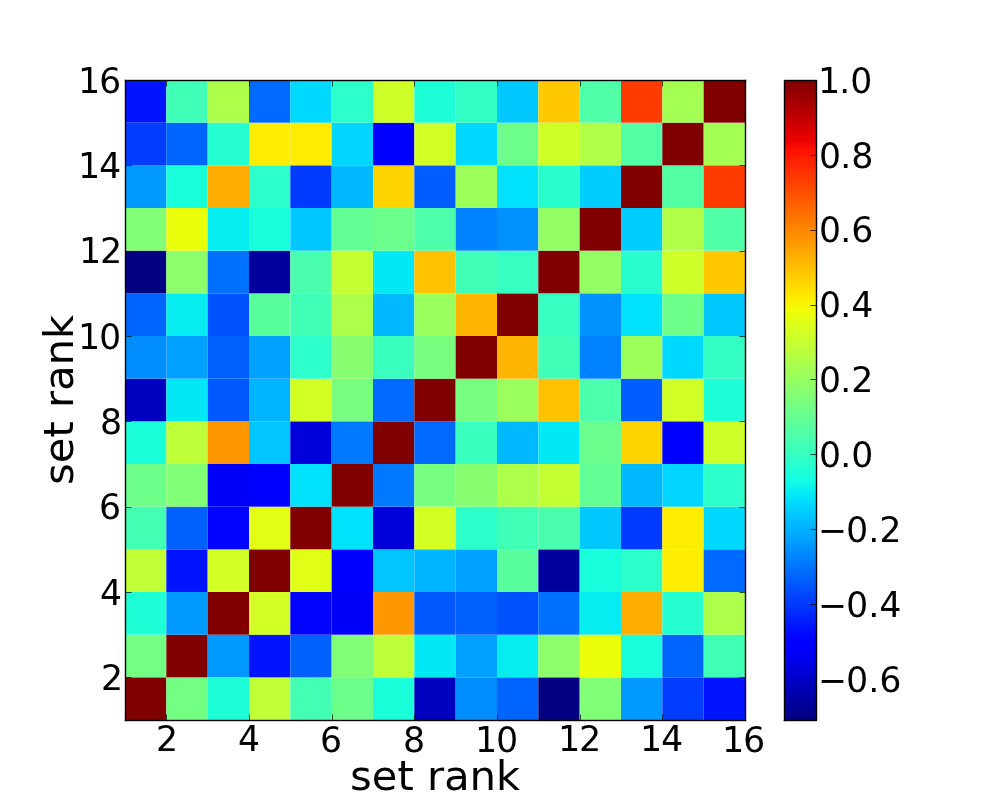}
\end{center}
\caption{\footnotesize (Color online) The overlap between the significance profiles of 16 randomly generated populations  of a thousand networks with initial connection probability $p_0=0.5$.  The color code is given on the side bar.   The projections of the SPs of randomized graphs on each other are scattered around zero, taking on negative as well  and positive values.  }
\label{fig:O3}
\end{figure}

As a final significance test, we have calculated the distribution of the numerical values of the overlaps (see Eq.\ref{eq:overlap}) amongst the SPs of the evolved sets (Fig.\ref{fig:O1}) and amongst the random sets (Fig.\ref{fig:O3}).  The distribution for the random SPs is symmetrical and more or less bell-shaped. The  mean overlap, taken between all pairs of random SPs is $\bar{\cal O}_{\rm ran} = -0.014$, the mode is at 0, the standard deviation $\bar{\sigma}_{O,{\rm ran}} = 0.28$ and the range is  $[-0.72, 0.72]$.  The mean overlap between all pairs of {\it evolved} SPs is $\bar{\cal O}_{\rm ev} = 0.46$, while the mode is at 0.6, and the standard deviation is $\bar{\sigma}_{O,{\rm ev}} = 0.31$. The distribution is skewed towards the higher values, except for a small tail at the lower edge, and has a range of $[-0.44, 0.92]$. Of  the evolved overlap  distribution,   73~\% lies beyond one  standard deviation of the null distribution,  43~\%  beyond two standard deviations and 10~\% beyond three standard deviations.  A major part of our evolved overlap distribution is therefore separated from the background at $p=0.5$ level of significance.


\end{document}